\documentclass[sigconf,nonacm]{acmart}

\usepackage{booktabs}
\usepackage{colortbl}
\usepackage[capitalise,noabbrev]{cleveref}
\newcommand{\pubCells}{224}

\newcommand{\pubCellsMovingPct}{21\%}

\newcommand{\pubPairs}{220}
\newcommand{\pubPairsReversed}{18}

\newcommand{\pubRunsReversingPct}{2.8\%}
\newcommand{\pubEmptyCells}{3}
\newcommand{\pubCleanCellsMovingPct}{20\%}

\newcommand{\pubScenarios}{6}

\newcommand{\pubOrderRuns}{29}
\newcommand{\pubOrderSame}{26}
\newcommand{\pubAllCells}{1023}
\newcommand{\pubAllCellsMovingPct}{22\%}

\newcommand{\pubTextCells}{101}
\newcommand{\pubTextQueries}{28}
\newcommand{\pubTextCellsMoving}{21}
\newcommand{\pubTextCellsMovingPct}{21\%}
\newcommand{\pubTextCellsBig}{5}
\newcommand{\pubTextMaxRange}{0.44}
\newcommand{\pubTextPairs}{112}
\newcommand{\pubTextPairsReversed}{7}

\newcommand{\pubTextMedicalCells}{11}
\newcommand{\pubTextMedicalMoving}{6}
\newcommand{\pubTextMmqaCells}{27}
\newcommand{\pubTextMmqaMoving}{1}

\newcommand{\flipGptoss}{10.2\%}
\newcommand{\flipCiGptoss}{[9.55\%, 10.85\%]}

\newcommand{\queriesGptoss}{10}

\newcommand{\flipMacroGptoss}{8.4\%}
\newcommand{\disagreeGptoss}{5.19\%}

\newcommand{\reasonAucGptoss}{0.84}
\newcommand{\reasonAucCiGptoss}{[0.83, 0.86]}
\newcommand{\reasonAucWithinGptoss}{0.79}
\newcommand{\reasonAucWithinMinGptoss}{0.73}
\newcommand{\reasonAucWithinMaxGptoss}{1.00}
\newcommand{\reasonAucQueriesGptoss}{6}
\newcommand{\reasonAucFlippedGptoss}{1{,}057}

\newcommand{\marginAucLlamaR}{1.00}

\newcommand{\marginFlippedLlamaR}{17}

\newcommand{\marginAucQwen}{1.00}

\newcommand{\marginFlippedQwen}{5}
\newcommand{\flipMistral}{0.07\%}

\newcommand{\flipMacroMistral}{0.03\%}

\newcommand{\disagreeGranite}{1.38\%}
\newcommand{\marginAucGranite}{0.93}
\newcommand{\marginAucCiGranite}{[0.90, 0.96]}
\newcommand{\marginFlippedGranite}{80}
\newcommand{\contestedFlipped}{77\%}
\newcommand{\contestedStable}{34\%}

\newcommand{\stackItemsMovie}{4{,}736}
\newcommand{\stackDifferMovie}{83}
\newcommand{\stackDifferPctMovie}{1.8\%}
\newcommand{\stackFlipMovie}{31}
\newcommand{\stackFlipPctMovie}{0.7\%}
\newcommand{\modelPairs}{126}
\newcommand{\modelPairsReversed}{22}

\newcommand{\topRunShare}{40\%}

\newcommand{\flipKTwo}{5.2\%}

\newcommand{\flipKFive}{10.2\%}
\newcommand{\flipKTen}{13.0\%}
\newcommand{\heldoutQueries}{10}
\newcommand{\heldoutWithin}{8}

\newcommand{\heldoutWorstPred}{1.00}
\newcommand{\heldoutWorstMeas}{0.30}

\newcommand{\dissentZeroGptoss}{3.2\%}

\newcommand{\dissentThreeGptoss}{37.5\%}
\newcommand{\dissentZeroLlamaW}{0.19\%}

\newcommand{\dissentThreeLlamaW}{10.7\%}

\newcommand{\flipBoolGptoss}{3.4\%}
\newcommand{\flipBoolLlamaW}{0.19\%}

\newcommand{\flipExtractGptoss}{24.0\%}
\newcommand{\flipExtractLlamaW}{7.8\%}

\newcommand{\flipLabelGptoss}{3.3\%}

\newcommand{\flipScoreGptoss}{17.5\%}

\newcommand{\voteThreeReduction}{36\%}
\newcommand{\voteFiveReduction}{48\%}
\newcommand{\minorityHeavy}{48\%}
\newcommand{\voteThreeMaxCost}{2.13}
\newcommand{\voteMaxAccuracyChange}{1.8}
\newcommand{\probeSettings}{6}

\newcommand{\probeConcOneDisagreeThree}{1.65\%}

\newcommand{\probeEffortLowMovedNine}{14.1\%}

\newcommand{\probeEffortMediumDisagreeNine}{8.82\%}

\newcommand{\probeEffortHighMovedNine}{10.9\%}

\newcommand{\casReproMedOneGptoss}{24\%}
\newcommand{\casSameMedOneGptoss}{1.0\%}
\newcommand{\casOtherMedOneGptoss}{2.2\%}

\newcommand{\casDisMedOneGptoss}{2.2\%}

\newcommand{\casItems}{1200}
\newcommand{\ampEveryCells}{17}
\newcommand{\ampEveryWithin}{16}
\newcommand{\ampEveryMae}{0.04}
\newcommand{\ampOtherCells}{36}
\newcommand{\ampOtherAbove}{32}
\newcommand{\flipKTwoScore}{8.8\%}
\newcommand{\flipKTenScore}{22.3\%}

\newcommand{\dupExampleDiffer}{6}
\newcommand{\dupExampleRuns}{10}
\newcommand{\dupAnyWithinGptoss}{91}
\newcommand{\dupInputs}{533}
\newcommand{\dupWithinGptoss}{6.0\%}
\newcommand{\dupAcrossGptoss}{6.1\%}
\newcommand{\dupWithinCiGptoss}{5.4--6.7\%}

\newcommand{\dupWithinGranite}{1.1\%}
\newcommand{\dupAcrossGranite}{1.3\%}

\newcommand{\selRemovedFull}{30\%}

\newcommand{\selOfFullTwenty}{64\%}
\newcommand{\selRandomOfFullTwenty}{29\%}
\newcommand{\selOfFullThirty}{80\%}

\newcommand{\bndLow}{9}
\newcommand{\bndHigh}{13}

\newcommand{\bndGold}{14}

\newcommand{\bndMedOneMissing}{1.6}
\newcommand{\bndMedOneOutside}{3.0}

\newcommand{\ladderUdaQualMin}{23\%}
\newcommand{\ladderUdaQualMax}{42\%}
\newcommand{\ladderSemOut}{32\%}
\newcommand{\ladderSemAns}{48\%}
\newcommand{\ladderSemQual}{52\%}
\newcommand{\ladderLroOut}{44\%}
\newcommand{\ladderLroAns}{59\%}
\newcommand{\ladderLroQual}{71\%}
\newcommand{\ladderUdaOut}{2\%}
\newcommand{\ladderUdaAns}{19\%}
\newcommand{\ladderUdaQual}{33\%}
\newcommand{\needPubPairs}{112}
\newcommand{\needPubMoreOne}{12}
\newcommand{\needPubMoreFive}{7}
\newcommand{\needPubMoreTen}{5}
\newcommand{\needPubMedian}{9}

\newcommand{\needDepMoreFive}{17}

\newcommand{\lroQueriesGptoss}{735}

\newcommand{\lroAnsweredGptoss}{726}
\newcommand{\lroDifferGptoss}{413}
\newcommand{\lroDifferPctGptoss}{57\%}
\newcommand{\lroMovesGptoss}{343}
\newcommand{\lroRangeGptoss}{0.107}

\newcommand{\lroFailLlamaR}{64}
\newcommand{\lroFailPctLlamaR}{9\%}

\newcommand{\lroDifferPctLlamaR}{25\%}

\newcommand{\lroRangeLlamaR}{0.026}

\newcommand{\lroQueriesLlamaW}{735}
\newcommand{\lroFailLlamaW}{437}
\newcommand{\lroFailPctLlamaW}{59\%}
\newcommand{\lroMixedFailLlamaW}{43}

\newcommand{\lroDifferPctGranite}{37\%}

\newcommand{\lroSettings}{14}
\newcommand{\lroSettingMinPctGptoss}{35\%}
\newcommand{\lroSettingMaxPctGptoss}{97\%}

\newcommand{\lroDifferPctClusterGptoss}{62\%}
\newcommand{\lroDifferPctImputeGptoss}{69\%}
\newcommand{\lroDifferPctMatchGptoss}{59\%}
\newcommand{\lroDifferPctMultiGptoss}{37\%}
\newcommand{\lroDifferPctOrderGptoss}{62\%}
\newcommand{\lroDifferPctSelectGptoss}{38\%}

\newcommand{\udaValueFlipPctGptoss}{7.5\%}
\newcommand{\udaNullShareGptoss}{31\%}
\newcommand{\udaValueToValueGptoss}{6.1\%}

\newcommand{\udaValuesGptoss}{2{,}208}

\newcommand{\udaDifferPctGptoss}{68\%}
\newcommand{\udaMovesGptoss}{78}
\newcommand{\udaScoredGptoss}{101}

\newcommand{\udaRangeGptoss}{0.031}
\newcommand{\udaMaxRangeGptoss}{0.17}

\newcommand{\udaDifferJoinGptoss}{13}

\newcommand{\udaValueFlipPctLlamaW}{4.0\%}

\newcommand{\udaValueToValueLlamaW}{3.8\%}

\newcommand{\udaDifferLlamaW}{83}
\newcommand{\udaDifferPctLlamaW}{67\%}

\newcommand{\udaValueToValueGranite}{4.5\%}

\newcommand{\udaQueries}{123}
\newcommand{\udaGoldEmpty}{22}

\newcommand{\prodFlipSentimentScore}{11\%}

\newcommand{\prodFlipSummary}{82\%}

\newcommand{\cellval}[4]{\csname cell@#1@#2@#3@#4\endcsname}
\expandafter\def\csname cell@medical@1@Granite@k\endcsname{5}
\expandafter\def\csname cell@medical@1@Granite@items\endcsname{1200}
\expandafter\def\csname cell@medical@1@Granite@flip\endcsname{2.7\%}
\expandafter\def\csname cell@medical@1@Granite@repro\endcsname{0.00}
\expandafter\def\csname cell@medical@1@Granite@indep\endcsname{0.00}
\expandafter\def\csname cell@medical@1@Granite@distinct\endcsname{5}
\expandafter\def\csname cell@medical@1@Granite@qmin\endcsname{0.34}
\expandafter\def\csname cell@medical@1@Granite@qmax\endcsname{0.38}
\expandafter\def\csname cell@medical@1@Granite@qrange\endcsname{0.04}
\expandafter\def\csname cell@medical@4@Granite@k\endcsname{5}
\expandafter\def\csname cell@medical@4@Granite@items\endcsname{1200}
\expandafter\def\csname cell@medical@4@Granite@flip\endcsname{0.1\%}
\expandafter\def\csname cell@medical@4@Granite@repro\endcsname{0.40}
\expandafter\def\csname cell@medical@4@Granite@indep\endcsname{0.54}
\expandafter\def\csname cell@medical@4@Granite@distinct\endcsname{2}
\expandafter\def\csname cell@medical@4@Granite@qmin\endcsname{1.00}
\expandafter\def\csname cell@medical@4@Granite@qmax\endcsname{1.00}
\expandafter\def\csname cell@medical@4@Granite@qrange\endcsname{0.00}
\expandafter\def\csname cell@medical@10@Granite@k\endcsname{5}
\expandafter\def\csname cell@medical@10@Granite@items\endcsname{1200}
\expandafter\def\csname cell@medical@10@Granite@flip\endcsname{3.8\%}
\expandafter\def\csname cell@medical@10@Granite@repro\endcsname{0.00}
\expandafter\def\csname cell@medical@10@Granite@indep\endcsname{0.00}
\expandafter\def\csname cell@medical@10@Granite@distinct\endcsname{5}
\expandafter\def\csname cell@medical@10@Granite@qmin\endcsname{0.22}
\expandafter\def\csname cell@medical@10@Granite@qmax\endcsname{0.23}
\expandafter\def\csname cell@medical@10@Granite@qrange\endcsname{0.01}
\expandafter\def\csname cell@medical@1@LlamaR@k\endcsname{5}
\expandafter\def\csname cell@medical@1@LlamaR@items\endcsname{1200}
\expandafter\def\csname cell@medical@1@LlamaR@flip\endcsname{0.0\%}
\expandafter\def\csname cell@medical@1@LlamaR@repro\endcsname{1.00}
\expandafter\def\csname cell@medical@1@LlamaR@indep\endcsname{1.00}
\expandafter\def\csname cell@medical@1@LlamaR@distinct\endcsname{1}
\expandafter\def\csname cell@medical@1@LlamaR@qmin\endcsname{0.28}
\expandafter\def\csname cell@medical@1@LlamaR@qmax\endcsname{0.28}
\expandafter\def\csname cell@medical@1@LlamaR@qrange\endcsname{0.00}
\expandafter\def\csname cell@medical@4@LlamaR@k\endcsname{5}
\expandafter\def\csname cell@medical@4@LlamaR@items\endcsname{1200}
\expandafter\def\csname cell@medical@4@LlamaR@flip\endcsname{0.1\%}
\expandafter\def\csname cell@medical@4@LlamaR@repro\endcsname{0.40}
\expandafter\def\csname cell@medical@4@LlamaR@indep\endcsname{0.52}
\expandafter\def\csname cell@medical@4@LlamaR@distinct\endcsname{2}
\expandafter\def\csname cell@medical@4@LlamaR@qmin\endcsname{0.98}
\expandafter\def\csname cell@medical@4@LlamaR@qmax\endcsname{0.98}
\expandafter\def\csname cell@medical@4@LlamaR@qrange\endcsname{0.00}
\expandafter\def\csname cell@medical@10@LlamaR@k\endcsname{5}
\expandafter\def\csname cell@medical@10@LlamaR@items\endcsname{1200}
\expandafter\def\csname cell@medical@10@LlamaR@flip\endcsname{0.7\%}
\expandafter\def\csname cell@medical@10@LlamaR@repro\endcsname{0.00}
\expandafter\def\csname cell@medical@10@LlamaR@indep\endcsname{0.02}
\expandafter\def\csname cell@medical@10@LlamaR@distinct\endcsname{5}
\expandafter\def\csname cell@medical@10@LlamaR@qmin\endcsname{0.36}
\expandafter\def\csname cell@medical@10@LlamaR@qmax\endcsname{0.36}
\expandafter\def\csname cell@medical@10@LlamaR@qrange\endcsname{0.00}
\expandafter\def\csname cell@medical@1@Qwen@k\endcsname{5}
\expandafter\def\csname cell@medical@1@Qwen@items\endcsname{1200}
\expandafter\def\csname cell@medical@1@Qwen@flip\endcsname{0.0\%}
\expandafter\def\csname cell@medical@1@Qwen@repro\endcsname{1.00}
\expandafter\def\csname cell@medical@1@Qwen@indep\endcsname{1.00}
\expandafter\def\csname cell@medical@1@Qwen@distinct\endcsname{1}
\expandafter\def\csname cell@medical@1@Qwen@qmin\endcsname{0.30}
\expandafter\def\csname cell@medical@1@Qwen@qmax\endcsname{0.30}
\expandafter\def\csname cell@medical@1@Qwen@qrange\endcsname{0.00}
\expandafter\def\csname cell@medical@4@Qwen@k\endcsname{5}
\expandafter\def\csname cell@medical@4@Qwen@items\endcsname{1200}
\expandafter\def\csname cell@medical@4@Qwen@flip\endcsname{0.1\%}
\expandafter\def\csname cell@medical@4@Qwen@repro\endcsname{0.60}
\expandafter\def\csname cell@medical@4@Qwen@indep\endcsname{0.67}
\expandafter\def\csname cell@medical@4@Qwen@distinct\endcsname{2}
\expandafter\def\csname cell@medical@4@Qwen@qmin\endcsname{0.99}
\expandafter\def\csname cell@medical@4@Qwen@qmax\endcsname{1.00}
\expandafter\def\csname cell@medical@4@Qwen@qrange\endcsname{0.01}
\expandafter\def\csname cell@medical@10@Qwen@k\endcsname{5}
\expandafter\def\csname cell@medical@10@Qwen@items\endcsname{1200}
\expandafter\def\csname cell@medical@10@Qwen@flip\endcsname{0.0\%}
\expandafter\def\csname cell@medical@10@Qwen@repro\endcsname{1.00}
\expandafter\def\csname cell@medical@10@Qwen@indep\endcsname{1.00}
\expandafter\def\csname cell@medical@10@Qwen@distinct\endcsname{1}
\expandafter\def\csname cell@medical@10@Qwen@qmin\endcsname{0.49}
\expandafter\def\csname cell@medical@10@Qwen@qmax\endcsname{0.49}
\expandafter\def\csname cell@medical@10@Qwen@qrange\endcsname{0.00}
\expandafter\def\csname cell@medical@1@Gptoss@k\endcsname{10}
\expandafter\def\csname cell@medical@1@Gptoss@items\endcsname{1200}
\expandafter\def\csname cell@medical@1@Gptoss@flip\endcsname{4.5\%}
\expandafter\def\csname cell@medical@1@Gptoss@repro\endcsname{0.00}
\expandafter\def\csname cell@medical@1@Gptoss@indep\endcsname{0.00}
\expandafter\def\csname cell@medical@1@Gptoss@distinct\endcsname{10}
\expandafter\def\csname cell@medical@1@Gptoss@qmin\endcsname{0.34}
\expandafter\def\csname cell@medical@1@Gptoss@qmax\endcsname{0.38}
\expandafter\def\csname cell@medical@1@Gptoss@qrange\endcsname{0.04}
\expandafter\def\csname cell@medical@4@Gptoss@k\endcsname{10}
\expandafter\def\csname cell@medical@4@Gptoss@items\endcsname{1200}
\expandafter\def\csname cell@medical@4@Gptoss@flip\endcsname{0.8\%}
\expandafter\def\csname cell@medical@4@Gptoss@repro\endcsname{0.00}
\expandafter\def\csname cell@medical@4@Gptoss@indep\endcsname{0.01}
\expandafter\def\csname cell@medical@4@Gptoss@distinct\endcsname{10}
\expandafter\def\csname cell@medical@4@Gptoss@qmin\endcsname{0.98}
\expandafter\def\csname cell@medical@4@Gptoss@qmax\endcsname{1.00}
\expandafter\def\csname cell@medical@4@Gptoss@qrange\endcsname{0.02}
\expandafter\def\csname cell@medical@10@Gptoss@k\endcsname{10}
\expandafter\def\csname cell@medical@10@Gptoss@items\endcsname{1200}
\expandafter\def\csname cell@medical@10@Gptoss@flip\endcsname{24.0\%}
\expandafter\def\csname cell@medical@10@Gptoss@repro\endcsname{0.00}
\expandafter\def\csname cell@medical@10@Gptoss@indep\endcsname{0.00}
\expandafter\def\csname cell@medical@10@Gptoss@distinct\endcsname{10}
\expandafter\def\csname cell@medical@10@Gptoss@qmin\endcsname{0.52}
\expandafter\def\csname cell@medical@10@Gptoss@qmax\endcsname{0.55}
\expandafter\def\csname cell@medical@10@Gptoss@qrange\endcsname{0.03}
\expandafter\def\csname cell@medical@1@LlamaW@k\endcsname{5}
\expandafter\def\csname cell@medical@1@LlamaW@items\endcsname{1200}
\expandafter\def\csname cell@medical@1@LlamaW@flip\endcsname{0.4\%}
\expandafter\def\csname cell@medical@1@LlamaW@repro\endcsname{0.00}
\expandafter\def\csname cell@medical@1@LlamaW@indep\endcsname{0.11}
\expandafter\def\csname cell@medical@1@LlamaW@distinct\endcsname{5}
\expandafter\def\csname cell@medical@1@LlamaW@qmin\endcsname{0.26}
\expandafter\def\csname cell@medical@1@LlamaW@qmax\endcsname{0.27}
\expandafter\def\csname cell@medical@1@LlamaW@qrange\endcsname{0.01}
\expandafter\def\csname cell@medical@4@LlamaW@k\endcsname{5}
\expandafter\def\csname cell@medical@4@LlamaW@items\endcsname{1200}
\expandafter\def\csname cell@medical@4@LlamaW@flip\endcsname{0.1\%}
\expandafter\def\csname cell@medical@4@LlamaW@repro\endcsname{0.60}
\expandafter\def\csname cell@medical@4@LlamaW@indep\endcsname{0.68}
\expandafter\def\csname cell@medical@4@LlamaW@distinct\endcsname{2}
\expandafter\def\csname cell@medical@4@LlamaW@qmin\endcsname{0.98}
\expandafter\def\csname cell@medical@4@LlamaW@qmax\endcsname{0.99}
\expandafter\def\csname cell@medical@4@LlamaW@qrange\endcsname{0.01}
\expandafter\def\csname cell@medical@10@LlamaW@k\endcsname{5}
\expandafter\def\csname cell@medical@10@LlamaW@items\endcsname{1200}
\expandafter\def\csname cell@medical@10@LlamaW@flip\endcsname{7.8\%}
\expandafter\def\csname cell@medical@10@LlamaW@repro\endcsname{0.00}
\expandafter\def\csname cell@medical@10@LlamaW@indep\endcsname{0.00}
\expandafter\def\csname cell@medical@10@LlamaW@distinct\endcsname{5}
\expandafter\def\csname cell@medical@10@LlamaW@qmin\endcsname{0.34}
\expandafter\def\csname cell@medical@10@LlamaW@qmax\endcsname{0.38}
\expandafter\def\csname cell@medical@10@LlamaW@qrange\endcsname{0.04}
\expandafter\def\csname cell@medical@1@Mistral@k\endcsname{5}
\expandafter\def\csname cell@medical@1@Mistral@items\endcsname{1200}
\expandafter\def\csname cell@medical@1@Mistral@flip\endcsname{0.0\%}
\expandafter\def\csname cell@medical@1@Mistral@repro\endcsname{1.00}
\expandafter\def\csname cell@medical@1@Mistral@indep\endcsname{1.00}
\expandafter\def\csname cell@medical@1@Mistral@distinct\endcsname{1}
\expandafter\def\csname cell@medical@1@Mistral@qmin\endcsname{0.48}
\expandafter\def\csname cell@medical@1@Mistral@qmax\endcsname{0.48}
\expandafter\def\csname cell@medical@1@Mistral@qrange\endcsname{0.00}
\expandafter\def\csname cell@medical@4@Mistral@k\endcsname{5}
\expandafter\def\csname cell@medical@4@Mistral@items\endcsname{1200}
\expandafter\def\csname cell@medical@4@Mistral@flip\endcsname{0.0\%}
\expandafter\def\csname cell@medical@4@Mistral@repro\endcsname{1.00}
\expandafter\def\csname cell@medical@4@Mistral@indep\endcsname{1.00}
\expandafter\def\csname cell@medical@4@Mistral@distinct\endcsname{1}
\expandafter\def\csname cell@medical@4@Mistral@qmin\endcsname{0.99}
\expandafter\def\csname cell@medical@4@Mistral@qmax\endcsname{0.99}
\expandafter\def\csname cell@medical@4@Mistral@qrange\endcsname{0.00}
\expandafter\def\csname cell@medical@10@Mistral@k\endcsname{5}
\expandafter\def\csname cell@medical@10@Mistral@items\endcsname{1200}
\expandafter\def\csname cell@medical@10@Mistral@flip\endcsname{0.1\%}
\expandafter\def\csname cell@medical@10@Mistral@repro\endcsname{0.60}
\expandafter\def\csname cell@medical@10@Mistral@indep\endcsname{0.69}
\expandafter\def\csname cell@medical@10@Mistral@distinct\endcsname{2}
\expandafter\def\csname cell@medical@10@Mistral@qmin\endcsname{0.44}
\expandafter\def\csname cell@medical@10@Mistral@qmax\endcsname{0.44}
\expandafter\def\csname cell@medical@10@Mistral@qrange\endcsname{0.00}
\expandafter\def\csname cell@movie@1@Granite@k\endcsname{5}
\expandafter\def\csname cell@movie@1@Granite@items\endcsname{2000}
\expandafter\def\csname cell@movie@1@Granite@flip\endcsname{3.2\%}
\expandafter\def\csname cell@movie@1@Granite@repro\endcsname{1.00}
\expandafter\def\csname cell@movie@1@Granite@indep\endcsname{1.00}
\expandafter\def\csname cell@movie@1@Granite@distinct\endcsname{1}
\expandafter\def\csname cell@movie@1@Granite@qmin\endcsname{1.00}
\expandafter\def\csname cell@movie@1@Granite@qmax\endcsname{1.00}
\expandafter\def\csname cell@movie@1@Granite@qrange\endcsname{0.00}
\expandafter\def\csname cell@movie@2@Granite@k\endcsname{5}
\expandafter\def\csname cell@movie@2@Granite@items\endcsname{120}
\expandafter\def\csname cell@movie@2@Granite@flip\endcsname{2.5\%}
\expandafter\def\csname cell@movie@2@Granite@repro\endcsname{1.00}
\expandafter\def\csname cell@movie@2@Granite@indep\endcsname{1.00}
\expandafter\def\csname cell@movie@2@Granite@distinct\endcsname{1}
\expandafter\def\csname cell@movie@2@Granite@qmin\endcsname{0.60}
\expandafter\def\csname cell@movie@2@Granite@qmax\endcsname{0.60}
\expandafter\def\csname cell@movie@2@Granite@qrange\endcsname{0.00}
\expandafter\def\csname cell@movie@3@Granite@k\endcsname{5}
\expandafter\def\csname cell@movie@3@Granite@items\endcsname{120}
\expandafter\def\csname cell@movie@3@Granite@flip\endcsname{2.5\%}
\expandafter\def\csname cell@movie@3@Granite@repro\endcsname{0.30}
\expandafter\def\csname cell@movie@3@Granite@indep\endcsname{0.32}
\expandafter\def\csname cell@movie@3@Granite@distinct\endcsname{3}
\expandafter\def\csname cell@movie@3@Granite@qmin\endcsname{0.64}
\expandafter\def\csname cell@movie@3@Granite@qmax\endcsname{0.79}
\expandafter\def\csname cell@movie@3@Granite@qrange\endcsname{0.14}
\expandafter\def\csname cell@movie@4@Granite@k\endcsname{5}
\expandafter\def\csname cell@movie@4@Granite@items\endcsname{120}
\expandafter\def\csname cell@movie@4@Granite@flip\endcsname{5.0\%}
\expandafter\def\csname cell@movie@4@Granite@repro\endcsname{0.10}
\expandafter\def\csname cell@movie@4@Granite@indep\endcsname{0.25}
\expandafter\def\csname cell@movie@4@Granite@distinct\endcsname{4}
\expandafter\def\csname cell@movie@4@Granite@qmin\endcsname{0.64}
\expandafter\def\csname cell@movie@4@Granite@qmax\endcsname{0.86}
\expandafter\def\csname cell@movie@4@Granite@qrange\endcsname{0.21}
\expandafter\def\csname cell@movie@8@Granite@k\endcsname{5}
\expandafter\def\csname cell@movie@8@Granite@items\endcsname{120}
\expandafter\def\csname cell@movie@8@Granite@flip\endcsname{1.7\%}
\expandafter\def\csname cell@movie@8@Granite@repro\endcsname{0.40}
\expandafter\def\csname cell@movie@8@Granite@indep\endcsname{0.51}
\expandafter\def\csname cell@movie@8@Granite@distinct\endcsname{2}
\expandafter\def\csname cell@movie@8@Granite@qmin\endcsname{0.80}
\expandafter\def\csname cell@movie@8@Granite@qmax\endcsname{0.84}
\expandafter\def\csname cell@movie@8@Granite@qrange\endcsname{0.04}
\expandafter\def\csname cell@movie@9@Granite@k\endcsname{5}
\expandafter\def\csname cell@movie@9@Granite@items\endcsname{256}
\expandafter\def\csname cell@movie@9@Granite@flip\endcsname{3.1\%}
\expandafter\def\csname cell@movie@9@Granite@repro\endcsname{0.00}
\expandafter\def\csname cell@movie@9@Granite@indep\endcsname{0.02}
\expandafter\def\csname cell@movie@9@Granite@distinct\endcsname{5}
\expandafter\def\csname cell@movie@9@Granite@qmin\endcsname{0.70}
\expandafter\def\csname cell@movie@9@Granite@qmax\endcsname{0.72}
\expandafter\def\csname cell@movie@9@Granite@qrange\endcsname{0.01}
\expandafter\def\csname cell@movie@10@Granite@k\endcsname{5}
\expandafter\def\csname cell@movie@10@Granite@items\endcsname{2000}
\expandafter\def\csname cell@movie@10@Granite@flip\endcsname{3.1\%}
\expandafter\def\csname cell@movie@10@Granite@repro\endcsname{0.00}
\expandafter\def\csname cell@movie@10@Granite@indep\endcsname{0.00}
\expandafter\def\csname cell@movie@10@Granite@distinct\endcsname{5}
\expandafter\def\csname cell@movie@10@Granite@qmin\endcsname{0.42}
\expandafter\def\csname cell@movie@10@Granite@qmax\endcsname{0.44}
\expandafter\def\csname cell@movie@10@Granite@qrange\endcsname{0.01}
\expandafter\def\csname cell@movie@5@Granite@k\endcsname{5}
\expandafter\def\csname cell@movie@5@Granite@items\endcsname{1560}
\expandafter\def\csname cell@movie@5@Granite@flip\endcsname{14.8\%}
\expandafter\def\csname cell@movie@5@Granite@repro\endcsname{0.00}
\expandafter\def\csname cell@movie@5@Granite@indep\endcsname{0.00}
\expandafter\def\csname cell@movie@5@Granite@distinct\endcsname{5}
\expandafter\def\csname cell@movie@5@Granite@qmin\endcsname{0.60}
\expandafter\def\csname cell@movie@5@Granite@qmax\endcsname{0.70}
\expandafter\def\csname cell@movie@5@Granite@qrange\endcsname{0.10}
\expandafter\def\csname cell@movie@6@Granite@k\endcsname{5}
\expandafter\def\csname cell@movie@6@Granite@items\endcsname{1560}
\expandafter\def\csname cell@movie@6@Granite@flip\endcsname{14.8\%}
\expandafter\def\csname cell@movie@6@Granite@repro\endcsname{0.00}
\expandafter\def\csname cell@movie@6@Granite@indep\endcsname{0.03}
\expandafter\def\csname cell@movie@6@Granite@distinct\endcsname{5}
\expandafter\def\csname cell@movie@6@Granite@qmin\endcsname{0.95}
\expandafter\def\csname cell@movie@6@Granite@qmax\endcsname{1.00}
\expandafter\def\csname cell@movie@6@Granite@qrange\endcsname{0.05}
\expandafter\def\csname cell@movie@7@Granite@k\endcsname{5}
\expandafter\def\csname cell@movie@7@Granite@items\endcsname{1560}
\expandafter\def\csname cell@movie@7@Granite@flip\endcsname{15.6\%}
\expandafter\def\csname cell@movie@7@Granite@repro\endcsname{0.00}
\expandafter\def\csname cell@movie@7@Granite@indep\endcsname{0.00}
\expandafter\def\csname cell@movie@7@Granite@distinct\endcsname{5}
\expandafter\def\csname cell@movie@7@Granite@qmin\endcsname{0.64}
\expandafter\def\csname cell@movie@7@Granite@qmax\endcsname{0.66}
\expandafter\def\csname cell@movie@7@Granite@qrange\endcsname{0.03}
\expandafter\def\csname cell@movie@1@LlamaR@k\endcsname{5}
\expandafter\def\csname cell@movie@1@LlamaR@items\endcsname{2000}
\expandafter\def\csname cell@movie@1@LlamaR@flip\endcsname{0.1\%}
\expandafter\def\csname cell@movie@1@LlamaR@repro\endcsname{1.00}
\expandafter\def\csname cell@movie@1@LlamaR@indep\endcsname{1.00}
\expandafter\def\csname cell@movie@1@LlamaR@distinct\endcsname{1}
\expandafter\def\csname cell@movie@1@LlamaR@qmin\endcsname{1.00}
\expandafter\def\csname cell@movie@1@LlamaR@qmax\endcsname{1.00}
\expandafter\def\csname cell@movie@1@LlamaR@qrange\endcsname{0.00}
\expandafter\def\csname cell@movie@2@LlamaR@k\endcsname{5}
\expandafter\def\csname cell@movie@2@LlamaR@items\endcsname{120}
\expandafter\def\csname cell@movie@2@LlamaR@flip\endcsname{0.0\%}
\expandafter\def\csname cell@movie@2@LlamaR@repro\endcsname{1.00}
\expandafter\def\csname cell@movie@2@LlamaR@indep\endcsname{1.00}
\expandafter\def\csname cell@movie@2@LlamaR@distinct\endcsname{1}
\expandafter\def\csname cell@movie@2@LlamaR@qmin\endcsname{1.00}
\expandafter\def\csname cell@movie@2@LlamaR@qmax\endcsname{1.00}
\expandafter\def\csname cell@movie@2@LlamaR@qrange\endcsname{0.00}
\expandafter\def\csname cell@movie@3@LlamaR@k\endcsname{5}
\expandafter\def\csname cell@movie@3@LlamaR@items\endcsname{120}
\expandafter\def\csname cell@movie@3@LlamaR@flip\endcsname{0.0\%}
\expandafter\def\csname cell@movie@3@LlamaR@repro\endcsname{1.00}
\expandafter\def\csname cell@movie@3@LlamaR@indep\endcsname{1.00}
\expandafter\def\csname cell@movie@3@LlamaR@distinct\endcsname{1}
\expandafter\def\csname cell@movie@3@LlamaR@qmin\endcsname{0.36}
\expandafter\def\csname cell@movie@3@LlamaR@qmax\endcsname{0.36}
\expandafter\def\csname cell@movie@3@LlamaR@qrange\endcsname{0.00}
\expandafter\def\csname cell@movie@4@LlamaR@k\endcsname{5}
\expandafter\def\csname cell@movie@4@LlamaR@items\endcsname{120}
\expandafter\def\csname cell@movie@4@LlamaR@flip\endcsname{0.0\%}
\expandafter\def\csname cell@movie@4@LlamaR@repro\endcsname{1.00}
\expandafter\def\csname cell@movie@4@LlamaR@indep\endcsname{1.00}
\expandafter\def\csname cell@movie@4@LlamaR@distinct\endcsname{1}
\expandafter\def\csname cell@movie@4@LlamaR@qmin\endcsname{0.36}
\expandafter\def\csname cell@movie@4@LlamaR@qmax\endcsname{0.36}
\expandafter\def\csname cell@movie@4@LlamaR@qrange\endcsname{0.00}
\expandafter\def\csname cell@movie@8@LlamaR@k\endcsname{5}
\expandafter\def\csname cell@movie@8@LlamaR@items\endcsname{120}
\expandafter\def\csname cell@movie@8@LlamaR@flip\endcsname{0.0\%}
\expandafter\def\csname cell@movie@8@LlamaR@repro\endcsname{1.00}
\expandafter\def\csname cell@movie@8@LlamaR@indep\endcsname{1.00}
\expandafter\def\csname cell@movie@8@LlamaR@distinct\endcsname{1}
\expandafter\def\csname cell@movie@8@LlamaR@qmin\endcsname{0.92}
\expandafter\def\csname cell@movie@8@LlamaR@qmax\endcsname{0.92}
\expandafter\def\csname cell@movie@8@LlamaR@qrange\endcsname{0.00}
\expandafter\def\csname cell@movie@9@LlamaR@k\endcsname{5}
\expandafter\def\csname cell@movie@9@LlamaR@items\endcsname{256}
\expandafter\def\csname cell@movie@9@LlamaR@flip\endcsname{0.8\%}
\expandafter\def\csname cell@movie@9@LlamaR@repro\endcsname{0.20}
\expandafter\def\csname cell@movie@9@LlamaR@indep\endcsname{0.35}
\expandafter\def\csname cell@movie@9@LlamaR@distinct\endcsname{3}
\expandafter\def\csname cell@movie@9@LlamaR@qmin\endcsname{0.76}
\expandafter\def\csname cell@movie@9@LlamaR@qmax\endcsname{0.77}
\expandafter\def\csname cell@movie@9@LlamaR@qrange\endcsname{0.00}
\expandafter\def\csname cell@movie@10@LlamaR@k\endcsname{5}
\expandafter\def\csname cell@movie@10@LlamaR@items\endcsname{2000}
\expandafter\def\csname cell@movie@10@LlamaR@flip\endcsname{0.6\%}
\expandafter\def\csname cell@movie@10@LlamaR@repro\endcsname{0.00}
\expandafter\def\csname cell@movie@10@LlamaR@indep\endcsname{0.03}
\expandafter\def\csname cell@movie@10@LlamaR@distinct\endcsname{5}
\expandafter\def\csname cell@movie@10@LlamaR@qmin\endcsname{0.38}
\expandafter\def\csname cell@movie@10@LlamaR@qmax\endcsname{0.39}
\expandafter\def\csname cell@movie@10@LlamaR@qrange\endcsname{0.01}
\expandafter\def\csname cell@movie@5@LlamaR@k\endcsname{5}
\expandafter\def\csname cell@movie@5@LlamaR@items\endcsname{1560}
\expandafter\def\csname cell@movie@5@LlamaR@flip\endcsname{0.3\%}
\expandafter\def\csname cell@movie@5@LlamaR@repro\endcsname{1.00}
\expandafter\def\csname cell@movie@5@LlamaR@indep\endcsname{1.00}
\expandafter\def\csname cell@movie@5@LlamaR@distinct\endcsname{1}
\expandafter\def\csname cell@movie@5@LlamaR@qmin\endcsname{0.90}
\expandafter\def\csname cell@movie@5@LlamaR@qmax\endcsname{0.90}
\expandafter\def\csname cell@movie@5@LlamaR@qrange\endcsname{0.00}
\expandafter\def\csname cell@movie@6@LlamaR@k\endcsname{5}
\expandafter\def\csname cell@movie@6@LlamaR@items\endcsname{1560}
\expandafter\def\csname cell@movie@6@LlamaR@flip\endcsname{0.2\%}
\expandafter\def\csname cell@movie@6@LlamaR@repro\endcsname{1.00}
\expandafter\def\csname cell@movie@6@LlamaR@indep\endcsname{1.00}
\expandafter\def\csname cell@movie@6@LlamaR@distinct\endcsname{1}
\expandafter\def\csname cell@movie@6@LlamaR@qmin\endcsname{1.00}
\expandafter\def\csname cell@movie@6@LlamaR@qmax\endcsname{1.00}
\expandafter\def\csname cell@movie@6@LlamaR@qrange\endcsname{0.00}
\expandafter\def\csname cell@movie@7@LlamaR@k\endcsname{5}
\expandafter\def\csname cell@movie@7@LlamaR@items\endcsname{1560}
\expandafter\def\csname cell@movie@7@LlamaR@flip\endcsname{0.2\%}
\expandafter\def\csname cell@movie@7@LlamaR@repro\endcsname{0.20}
\expandafter\def\csname cell@movie@7@LlamaR@indep\endcsname{0.15}
\expandafter\def\csname cell@movie@7@LlamaR@distinct\endcsname{3}
\expandafter\def\csname cell@movie@7@LlamaR@qmin\endcsname{0.63}
\expandafter\def\csname cell@movie@7@LlamaR@qmax\endcsname{0.63}
\expandafter\def\csname cell@movie@7@LlamaR@qrange\endcsname{0.00}
\expandafter\def\csname cell@movie@1@Qwen@k\endcsname{5}
\expandafter\def\csname cell@movie@1@Qwen@items\endcsname{2000}
\expandafter\def\csname cell@movie@1@Qwen@flip\endcsname{0.1\%}
\expandafter\def\csname cell@movie@1@Qwen@repro\endcsname{1.00}
\expandafter\def\csname cell@movie@1@Qwen@indep\endcsname{1.00}
\expandafter\def\csname cell@movie@1@Qwen@distinct\endcsname{1}
\expandafter\def\csname cell@movie@1@Qwen@qmin\endcsname{1.00}
\expandafter\def\csname cell@movie@1@Qwen@qmax\endcsname{1.00}
\expandafter\def\csname cell@movie@1@Qwen@qrange\endcsname{0.00}
\expandafter\def\csname cell@movie@2@Qwen@k\endcsname{5}
\expandafter\def\csname cell@movie@2@Qwen@items\endcsname{120}
\expandafter\def\csname cell@movie@2@Qwen@flip\endcsname{0.0\%}
\expandafter\def\csname cell@movie@2@Qwen@repro\endcsname{1.00}
\expandafter\def\csname cell@movie@2@Qwen@indep\endcsname{1.00}
\expandafter\def\csname cell@movie@2@Qwen@distinct\endcsname{1}
\expandafter\def\csname cell@movie@2@Qwen@qmin\endcsname{1.00}
\expandafter\def\csname cell@movie@2@Qwen@qmax\endcsname{1.00}
\expandafter\def\csname cell@movie@2@Qwen@qrange\endcsname{0.00}
\expandafter\def\csname cell@movie@3@Qwen@k\endcsname{5}
\expandafter\def\csname cell@movie@3@Qwen@items\endcsname{120}
\expandafter\def\csname cell@movie@3@Qwen@flip\endcsname{0.0\%}
\expandafter\def\csname cell@movie@3@Qwen@repro\endcsname{1.00}
\expandafter\def\csname cell@movie@3@Qwen@indep\endcsname{1.00}
\expandafter\def\csname cell@movie@3@Qwen@distinct\endcsname{1}
\expandafter\def\csname cell@movie@3@Qwen@qmin\endcsname{0.50}
\expandafter\def\csname cell@movie@3@Qwen@qmax\endcsname{0.50}
\expandafter\def\csname cell@movie@3@Qwen@qrange\endcsname{0.00}
\expandafter\def\csname cell@movie@4@Qwen@k\endcsname{5}
\expandafter\def\csname cell@movie@4@Qwen@items\endcsname{120}
\expandafter\def\csname cell@movie@4@Qwen@flip\endcsname{0.0\%}
\expandafter\def\csname cell@movie@4@Qwen@repro\endcsname{1.00}
\expandafter\def\csname cell@movie@4@Qwen@indep\endcsname{1.00}
\expandafter\def\csname cell@movie@4@Qwen@distinct\endcsname{1}
\expandafter\def\csname cell@movie@4@Qwen@qmin\endcsname{0.50}
\expandafter\def\csname cell@movie@4@Qwen@qmax\endcsname{0.50}
\expandafter\def\csname cell@movie@4@Qwen@qrange\endcsname{0.00}
\expandafter\def\csname cell@movie@8@Qwen@k\endcsname{5}
\expandafter\def\csname cell@movie@8@Qwen@items\endcsname{120}
\expandafter\def\csname cell@movie@8@Qwen@flip\endcsname{0.0\%}
\expandafter\def\csname cell@movie@8@Qwen@repro\endcsname{1.00}
\expandafter\def\csname cell@movie@8@Qwen@indep\endcsname{1.00}
\expandafter\def\csname cell@movie@8@Qwen@distinct\endcsname{1}
\expandafter\def\csname cell@movie@8@Qwen@qmin\endcsname{0.96}
\expandafter\def\csname cell@movie@8@Qwen@qmax\endcsname{0.96}
\expandafter\def\csname cell@movie@8@Qwen@qrange\endcsname{0.00}
\expandafter\def\csname cell@movie@9@Qwen@k\endcsname{5}
\expandafter\def\csname cell@movie@9@Qwen@items\endcsname{256}
\expandafter\def\csname cell@movie@9@Qwen@flip\endcsname{0.4\%}
\expandafter\def\csname cell@movie@9@Qwen@repro\endcsname{0.40}
\expandafter\def\csname cell@movie@9@Qwen@indep\endcsname{0.52}
\expandafter\def\csname cell@movie@9@Qwen@distinct\endcsname{2}
\expandafter\def\csname cell@movie@9@Qwen@qmin\endcsname{0.75}
\expandafter\def\csname cell@movie@9@Qwen@qmax\endcsname{0.76}
\expandafter\def\csname cell@movie@9@Qwen@qrange\endcsname{0.01}
\expandafter\def\csname cell@movie@10@Qwen@k\endcsname{5}
\expandafter\def\csname cell@movie@10@Qwen@items\endcsname{2000}
\expandafter\def\csname cell@movie@10@Qwen@flip\endcsname{0.1\%}
\expandafter\def\csname cell@movie@10@Qwen@repro\endcsname{1.00}
\expandafter\def\csname cell@movie@10@Qwen@indep\endcsname{1.00}
\expandafter\def\csname cell@movie@10@Qwen@distinct\endcsname{1}
\expandafter\def\csname cell@movie@10@Qwen@qmin\endcsname{0.42}
\expandafter\def\csname cell@movie@10@Qwen@qmax\endcsname{0.42}
\expandafter\def\csname cell@movie@10@Qwen@qrange\endcsname{0.00}
\expandafter\def\csname cell@movie@5@Qwen@k\endcsname{5}
\expandafter\def\csname cell@movie@5@Qwen@items\endcsname{1560}
\expandafter\def\csname cell@movie@5@Qwen@flip\endcsname{0.1\%}
\expandafter\def\csname cell@movie@5@Qwen@repro\endcsname{1.00}
\expandafter\def\csname cell@movie@5@Qwen@indep\endcsname{1.00}
\expandafter\def\csname cell@movie@5@Qwen@distinct\endcsname{1}
\expandafter\def\csname cell@movie@5@Qwen@qmin\endcsname{0.80}
\expandafter\def\csname cell@movie@5@Qwen@qmax\endcsname{0.80}
\expandafter\def\csname cell@movie@5@Qwen@qrange\endcsname{0.00}
\expandafter\def\csname cell@movie@6@Qwen@k\endcsname{5}
\expandafter\def\csname cell@movie@6@Qwen@items\endcsname{1560}
\expandafter\def\csname cell@movie@6@Qwen@flip\endcsname{0.2\%}
\expandafter\def\csname cell@movie@6@Qwen@repro\endcsname{1.00}
\expandafter\def\csname cell@movie@6@Qwen@indep\endcsname{1.00}
\expandafter\def\csname cell@movie@6@Qwen@distinct\endcsname{1}
\expandafter\def\csname cell@movie@6@Qwen@qmin\endcsname{0.70}
\expandafter\def\csname cell@movie@6@Qwen@qmax\endcsname{0.70}
\expandafter\def\csname cell@movie@6@Qwen@qrange\endcsname{0.00}
\expandafter\def\csname cell@movie@7@Qwen@k\endcsname{5}
\expandafter\def\csname cell@movie@7@Qwen@items\endcsname{1560}
\expandafter\def\csname cell@movie@7@Qwen@flip\endcsname{0.3\%}
\expandafter\def\csname cell@movie@7@Qwen@repro\endcsname{0.10}
\expandafter\def\csname cell@movie@7@Qwen@indep\endcsname{0.16}
\expandafter\def\csname cell@movie@7@Qwen@distinct\endcsname{4}
\expandafter\def\csname cell@movie@7@Qwen@qmin\endcsname{0.72}
\expandafter\def\csname cell@movie@7@Qwen@qmax\endcsname{0.72}
\expandafter\def\csname cell@movie@7@Qwen@qrange\endcsname{0.00}
\expandafter\def\csname cell@movie@1@Gptoss@k\endcsname{10}
\expandafter\def\csname cell@movie@1@Gptoss@items\endcsname{2000}
\expandafter\def\csname cell@movie@1@Gptoss@flip\endcsname{4.2\%}
\expandafter\def\csname cell@movie@1@Gptoss@repro\endcsname{0.53}
\expandafter\def\csname cell@movie@1@Gptoss@indep\endcsname{0.59}
\expandafter\def\csname cell@movie@1@Gptoss@distinct\endcsname{2}
\expandafter\def\csname cell@movie@1@Gptoss@qmin\endcsname{1.00}
\expandafter\def\csname cell@movie@1@Gptoss@qmax\endcsname{1.00}
\expandafter\def\csname cell@movie@1@Gptoss@qrange\endcsname{0.00}
\expandafter\def\csname cell@movie@2@Gptoss@k\endcsname{10}
\expandafter\def\csname cell@movie@2@Gptoss@items\endcsname{120}
\expandafter\def\csname cell@movie@2@Gptoss@flip\endcsname{4.4\%}
\expandafter\def\csname cell@movie@2@Gptoss@repro\endcsname{0.62}
\expandafter\def\csname cell@movie@2@Gptoss@indep\endcsname{0.37}
\expandafter\def\csname cell@movie@2@Gptoss@distinct\endcsname{3}
\expandafter\def\csname cell@movie@2@Gptoss@qmin\endcsname{0.40}
\expandafter\def\csname cell@movie@2@Gptoss@qmax\endcsname{0.60}
\expandafter\def\csname cell@movie@2@Gptoss@qrange\endcsname{0.20}
\expandafter\def\csname cell@movie@3@Gptoss@k\endcsname{10}
\expandafter\def\csname cell@movie@3@Gptoss@items\endcsname{120}
\expandafter\def\csname cell@movie@3@Gptoss@flip\endcsname{2.8\%}
\expandafter\def\csname cell@movie@3@Gptoss@repro\endcsname{0.36}
\expandafter\def\csname cell@movie@3@Gptoss@indep\endcsname{0.36}
\expandafter\def\csname cell@movie@3@Gptoss@distinct\endcsname{4}
\expandafter\def\csname cell@movie@3@Gptoss@qmin\endcsname{0.64}
\expandafter\def\csname cell@movie@3@Gptoss@qmax\endcsname{0.93}
\expandafter\def\csname cell@movie@3@Gptoss@qrange\endcsname{0.29}
\expandafter\def\csname cell@movie@4@Gptoss@k\endcsname{10}
\expandafter\def\csname cell@movie@4@Gptoss@items\endcsname{120}
\expandafter\def\csname cell@movie@4@Gptoss@flip\endcsname{3.1\%}
\expandafter\def\csname cell@movie@4@Gptoss@repro\endcsname{0.27}
\expandafter\def\csname cell@movie@4@Gptoss@indep\endcsname{0.32}
\expandafter\def\csname cell@movie@4@Gptoss@distinct\endcsname{4}
\expandafter\def\csname cell@movie@4@Gptoss@qmin\endcsname{0.64}
\expandafter\def\csname cell@movie@4@Gptoss@qmax\endcsname{0.86}
\expandafter\def\csname cell@movie@4@Gptoss@qrange\endcsname{0.21}
\expandafter\def\csname cell@movie@8@Gptoss@k\endcsname{10}
\expandafter\def\csname cell@movie@8@Gptoss@items\endcsname{120}
\expandafter\def\csname cell@movie@8@Gptoss@flip\endcsname{3.3\%}
\expandafter\def\csname cell@movie@8@Gptoss@repro\endcsname{0.29}
\expandafter\def\csname cell@movie@8@Gptoss@indep\endcsname{0.30}
\expandafter\def\csname cell@movie@8@Gptoss@distinct\endcsname{4}
\expandafter\def\csname cell@movie@8@Gptoss@qmin\endcsname{0.92}
\expandafter\def\csname cell@movie@8@Gptoss@qmax\endcsname{1.00}
\expandafter\def\csname cell@movie@8@Gptoss@qrange\endcsname{0.08}
\expandafter\def\csname cell@movie@9@Gptoss@k\endcsname{10}
\expandafter\def\csname cell@movie@9@Gptoss@items\endcsname{256}
\expandafter\def\csname cell@movie@9@Gptoss@flip\endcsname{19.4\%}
\expandafter\def\csname cell@movie@9@Gptoss@repro\endcsname{0.00}
\expandafter\def\csname cell@movie@9@Gptoss@indep\endcsname{0.00}
\expandafter\def\csname cell@movie@9@Gptoss@distinct\endcsname{10}
\expandafter\def\csname cell@movie@9@Gptoss@qmin\endcsname{0.74}
\expandafter\def\csname cell@movie@9@Gptoss@qmax\endcsname{0.77}
\expandafter\def\csname cell@movie@9@Gptoss@qrange\endcsname{0.03}
\expandafter\def\csname cell@movie@10@Gptoss@k\endcsname{10}
\expandafter\def\csname cell@movie@10@Gptoss@items\endcsname{2000}
\expandafter\def\csname cell@movie@10@Gptoss@flip\endcsname{17.3\%}
\expandafter\def\csname cell@movie@10@Gptoss@repro\endcsname{0.00}
\expandafter\def\csname cell@movie@10@Gptoss@indep\endcsname{0.00}
\expandafter\def\csname cell@movie@10@Gptoss@distinct\endcsname{10}
\expandafter\def\csname cell@movie@10@Gptoss@qmin\endcsname{0.41}
\expandafter\def\csname cell@movie@10@Gptoss@qmax\endcsname{0.43}
\expandafter\def\csname cell@movie@10@Gptoss@qrange\endcsname{0.03}
\expandafter\def\csname cell@movie@5@Gptoss@k\endcsname{10}
\expandafter\def\csname cell@movie@5@Gptoss@items\endcsname{1560}
\expandafter\def\csname cell@movie@5@Gptoss@flip\endcsname{13.5\%}
\expandafter\def\csname cell@movie@5@Gptoss@repro\endcsname{0.00}
\expandafter\def\csname cell@movie@5@Gptoss@indep\endcsname{0.00}
\expandafter\def\csname cell@movie@5@Gptoss@distinct\endcsname{10}
\expandafter\def\csname cell@movie@5@Gptoss@qmin\endcsname{0.20}
\expandafter\def\csname cell@movie@5@Gptoss@qmax\endcsname{0.70}
\expandafter\def\csname cell@movie@5@Gptoss@qrange\endcsname{0.50}
\expandafter\def\csname cell@movie@6@Gptoss@k\endcsname{10}
\expandafter\def\csname cell@movie@6@Gptoss@items\endcsname{1560}
\expandafter\def\csname cell@movie@6@Gptoss@flip\endcsname{15.1\%}
\expandafter\def\csname cell@movie@6@Gptoss@repro\endcsname{0.00}
\expandafter\def\csname cell@movie@6@Gptoss@indep\endcsname{0.01}
\expandafter\def\csname cell@movie@6@Gptoss@distinct\endcsname{10}
\expandafter\def\csname cell@movie@6@Gptoss@qmin\endcsname{0.50}
\expandafter\def\csname cell@movie@6@Gptoss@qmax\endcsname{0.80}
\expandafter\def\csname cell@movie@6@Gptoss@qrange\endcsname{0.30}
\expandafter\def\csname cell@movie@7@Gptoss@k\endcsname{10}
\expandafter\def\csname cell@movie@7@Gptoss@items\endcsname{1560}
\expandafter\def\csname cell@movie@7@Gptoss@flip\endcsname{15.2\%}
\expandafter\def\csname cell@movie@7@Gptoss@repro\endcsname{0.00}
\expandafter\def\csname cell@movie@7@Gptoss@indep\endcsname{0.00}
\expandafter\def\csname cell@movie@7@Gptoss@distinct\endcsname{10}
\expandafter\def\csname cell@movie@7@Gptoss@qmin\endcsname{0.69}
\expandafter\def\csname cell@movie@7@Gptoss@qmax\endcsname{0.71}
\expandafter\def\csname cell@movie@7@Gptoss@qrange\endcsname{0.02}
\expandafter\def\csname cell@movie@1@LlamaW@k\endcsname{5}
\expandafter\def\csname cell@movie@1@LlamaW@items\endcsname{2000}
\expandafter\def\csname cell@movie@1@LlamaW@flip\endcsname{0.1\%}
\expandafter\def\csname cell@movie@1@LlamaW@repro\endcsname{1.00}
\expandafter\def\csname cell@movie@1@LlamaW@indep\endcsname{1.00}
\expandafter\def\csname cell@movie@1@LlamaW@distinct\endcsname{1}
\expandafter\def\csname cell@movie@1@LlamaW@qmin\endcsname{1.00}
\expandafter\def\csname cell@movie@1@LlamaW@qmax\endcsname{1.00}
\expandafter\def\csname cell@movie@1@LlamaW@qrange\endcsname{0.00}
\expandafter\def\csname cell@movie@2@LlamaW@k\endcsname{5}
\expandafter\def\csname cell@movie@2@LlamaW@items\endcsname{120}
\expandafter\def\csname cell@movie@2@LlamaW@flip\endcsname{0.0\%}
\expandafter\def\csname cell@movie@2@LlamaW@repro\endcsname{1.00}
\expandafter\def\csname cell@movie@2@LlamaW@indep\endcsname{1.00}
\expandafter\def\csname cell@movie@2@LlamaW@distinct\endcsname{1}
\expandafter\def\csname cell@movie@2@LlamaW@qmin\endcsname{1.00}
\expandafter\def\csname cell@movie@2@LlamaW@qmax\endcsname{1.00}
\expandafter\def\csname cell@movie@2@LlamaW@qrange\endcsname{0.00}
\expandafter\def\csname cell@movie@3@LlamaW@k\endcsname{5}
\expandafter\def\csname cell@movie@3@LlamaW@items\endcsname{120}
\expandafter\def\csname cell@movie@3@LlamaW@flip\endcsname{0.0\%}
\expandafter\def\csname cell@movie@3@LlamaW@repro\endcsname{1.00}
\expandafter\def\csname cell@movie@3@LlamaW@indep\endcsname{1.00}
\expandafter\def\csname cell@movie@3@LlamaW@distinct\endcsname{1}
\expandafter\def\csname cell@movie@3@LlamaW@qmin\endcsname{0.43}
\expandafter\def\csname cell@movie@3@LlamaW@qmax\endcsname{0.43}
\expandafter\def\csname cell@movie@3@LlamaW@qrange\endcsname{0.00}
\expandafter\def\csname cell@movie@4@LlamaW@k\endcsname{5}
\expandafter\def\csname cell@movie@4@LlamaW@items\endcsname{120}
\expandafter\def\csname cell@movie@4@LlamaW@flip\endcsname{0.0\%}
\expandafter\def\csname cell@movie@4@LlamaW@repro\endcsname{1.00}
\expandafter\def\csname cell@movie@4@LlamaW@indep\endcsname{1.00}
\expandafter\def\csname cell@movie@4@LlamaW@distinct\endcsname{1}
\expandafter\def\csname cell@movie@4@LlamaW@qmin\endcsname{0.43}
\expandafter\def\csname cell@movie@4@LlamaW@qmax\endcsname{0.43}
\expandafter\def\csname cell@movie@4@LlamaW@qrange\endcsname{0.00}
\expandafter\def\csname cell@movie@8@LlamaW@k\endcsname{5}
\expandafter\def\csname cell@movie@8@LlamaW@items\endcsname{120}
\expandafter\def\csname cell@movie@8@LlamaW@flip\endcsname{0.0\%}
\expandafter\def\csname cell@movie@8@LlamaW@repro\endcsname{1.00}
\expandafter\def\csname cell@movie@8@LlamaW@indep\endcsname{1.00}
\expandafter\def\csname cell@movie@8@LlamaW@distinct\endcsname{1}
\expandafter\def\csname cell@movie@8@LlamaW@qmin\endcsname{0.96}
\expandafter\def\csname cell@movie@8@LlamaW@qmax\endcsname{0.96}
\expandafter\def\csname cell@movie@8@LlamaW@qrange\endcsname{0.00}
\expandafter\def\csname cell@movie@9@LlamaW@k\endcsname{5}
\expandafter\def\csname cell@movie@9@LlamaW@items\endcsname{256}
\expandafter\def\csname cell@movie@9@LlamaW@flip\endcsname{0.0\%}
\expandafter\def\csname cell@movie@9@LlamaW@repro\endcsname{1.00}
\expandafter\def\csname cell@movie@9@LlamaW@indep\endcsname{1.00}
\expandafter\def\csname cell@movie@9@LlamaW@distinct\endcsname{1}
\expandafter\def\csname cell@movie@9@LlamaW@qmin\endcsname{0.77}
\expandafter\def\csname cell@movie@9@LlamaW@qmax\endcsname{0.77}
\expandafter\def\csname cell@movie@9@LlamaW@qrange\endcsname{0.00}
\expandafter\def\csname cell@movie@10@LlamaW@k\endcsname{5}
\expandafter\def\csname cell@movie@10@LlamaW@items\endcsname{2000}
\expandafter\def\csname cell@movie@10@LlamaW@flip\endcsname{0.7\%}
\expandafter\def\csname cell@movie@10@LlamaW@repro\endcsname{0.10}
\expandafter\def\csname cell@movie@10@LlamaW@indep\endcsname{0.14}
\expandafter\def\csname cell@movie@10@LlamaW@distinct\endcsname{4}
\expandafter\def\csname cell@movie@10@LlamaW@qmin\endcsname{0.38}
\expandafter\def\csname cell@movie@10@LlamaW@qmax\endcsname{0.38}
\expandafter\def\csname cell@movie@10@LlamaW@qrange\endcsname{0.00}
\expandafter\def\csname cell@movie@5@LlamaW@k\endcsname{5}
\expandafter\def\csname cell@movie@5@LlamaW@items\endcsname{1560}
\expandafter\def\csname cell@movie@5@LlamaW@flip\endcsname{2.7\%}
\expandafter\def\csname cell@movie@5@LlamaW@repro\endcsname{0.10}
\expandafter\def\csname cell@movie@5@LlamaW@indep\endcsname{0.09}
\expandafter\def\csname cell@movie@5@LlamaW@distinct\endcsname{4}
\expandafter\def\csname cell@movie@5@LlamaW@qmin\endcsname{0.90}
\expandafter\def\csname cell@movie@5@LlamaW@qmax\endcsname{0.90}
\expandafter\def\csname cell@movie@5@LlamaW@qrange\endcsname{0.00}
\expandafter\def\csname cell@movie@6@LlamaW@k\endcsname{5}
\expandafter\def\csname cell@movie@6@LlamaW@items\endcsname{1560}
\expandafter\def\csname cell@movie@6@LlamaW@flip\endcsname{1.4\%}
\expandafter\def\csname cell@movie@6@LlamaW@repro\endcsname{0.30}
\expandafter\def\csname cell@movie@6@LlamaW@indep\endcsname{0.47}
\expandafter\def\csname cell@movie@6@LlamaW@distinct\endcsname{3}
\expandafter\def\csname cell@movie@6@LlamaW@qmin\endcsname{0.90}
\expandafter\def\csname cell@movie@6@LlamaW@qmax\endcsname{0.90}
\expandafter\def\csname cell@movie@6@LlamaW@qrange\endcsname{0.00}
\expandafter\def\csname cell@movie@7@LlamaW@k\endcsname{5}
\expandafter\def\csname cell@movie@7@LlamaW@items\endcsname{1560}
\expandafter\def\csname cell@movie@7@LlamaW@flip\endcsname{1.5\%}
\expandafter\def\csname cell@movie@7@LlamaW@repro\endcsname{0.00}
\expandafter\def\csname cell@movie@7@LlamaW@indep\endcsname{0.00}
\expandafter\def\csname cell@movie@7@LlamaW@distinct\endcsname{5}
\expandafter\def\csname cell@movie@7@LlamaW@qmin\endcsname{0.63}
\expandafter\def\csname cell@movie@7@LlamaW@qmax\endcsname{0.64}
\expandafter\def\csname cell@movie@7@LlamaW@qrange\endcsname{0.01}
\expandafter\def\csname cell@movie@1@Mistral@k\endcsname{5}
\expandafter\def\csname cell@movie@1@Mistral@items\endcsname{2000}
\expandafter\def\csname cell@movie@1@Mistral@flip\endcsname{0.1\%}
\expandafter\def\csname cell@movie@1@Mistral@repro\endcsname{1.00}
\expandafter\def\csname cell@movie@1@Mistral@indep\endcsname{1.00}
\expandafter\def\csname cell@movie@1@Mistral@distinct\endcsname{1}
\expandafter\def\csname cell@movie@1@Mistral@qmin\endcsname{1.00}
\expandafter\def\csname cell@movie@1@Mistral@qmax\endcsname{1.00}
\expandafter\def\csname cell@movie@1@Mistral@qrange\endcsname{0.00}
\expandafter\def\csname cell@movie@2@Mistral@k\endcsname{5}
\expandafter\def\csname cell@movie@2@Mistral@items\endcsname{120}
\expandafter\def\csname cell@movie@2@Mistral@flip\endcsname{0.0\%}
\expandafter\def\csname cell@movie@2@Mistral@repro\endcsname{1.00}
\expandafter\def\csname cell@movie@2@Mistral@indep\endcsname{1.00}
\expandafter\def\csname cell@movie@2@Mistral@distinct\endcsname{1}
\expandafter\def\csname cell@movie@2@Mistral@qmin\endcsname{1.00}
\expandafter\def\csname cell@movie@2@Mistral@qmax\endcsname{1.00}
\expandafter\def\csname cell@movie@2@Mistral@qrange\endcsname{0.00}
\expandafter\def\csname cell@movie@3@Mistral@k\endcsname{5}
\expandafter\def\csname cell@movie@3@Mistral@items\endcsname{120}
\expandafter\def\csname cell@movie@3@Mistral@flip\endcsname{0.0\%}
\expandafter\def\csname cell@movie@3@Mistral@repro\endcsname{1.00}
\expandafter\def\csname cell@movie@3@Mistral@indep\endcsname{1.00}
\expandafter\def\csname cell@movie@3@Mistral@distinct\endcsname{1}
\expandafter\def\csname cell@movie@3@Mistral@qmin\endcsname{0.36}
\expandafter\def\csname cell@movie@3@Mistral@qmax\endcsname{0.36}
\expandafter\def\csname cell@movie@3@Mistral@qrange\endcsname{0.00}
\expandafter\def\csname cell@movie@4@Mistral@k\endcsname{5}
\expandafter\def\csname cell@movie@4@Mistral@items\endcsname{120}
\expandafter\def\csname cell@movie@4@Mistral@flip\endcsname{0.0\%}
\expandafter\def\csname cell@movie@4@Mistral@repro\endcsname{1.00}
\expandafter\def\csname cell@movie@4@Mistral@indep\endcsname{1.00}
\expandafter\def\csname cell@movie@4@Mistral@distinct\endcsname{1}
\expandafter\def\csname cell@movie@4@Mistral@qmin\endcsname{0.36}
\expandafter\def\csname cell@movie@4@Mistral@qmax\endcsname{0.36}
\expandafter\def\csname cell@movie@4@Mistral@qrange\endcsname{0.00}
\expandafter\def\csname cell@movie@8@Mistral@k\endcsname{5}
\expandafter\def\csname cell@movie@8@Mistral@items\endcsname{120}
\expandafter\def\csname cell@movie@8@Mistral@flip\endcsname{0.0\%}
\expandafter\def\csname cell@movie@8@Mistral@repro\endcsname{1.00}
\expandafter\def\csname cell@movie@8@Mistral@indep\endcsname{1.00}
\expandafter\def\csname cell@movie@8@Mistral@distinct\endcsname{1}
\expandafter\def\csname cell@movie@8@Mistral@qmin\endcsname{0.80}
\expandafter\def\csname cell@movie@8@Mistral@qmax\endcsname{0.80}
\expandafter\def\csname cell@movie@8@Mistral@qrange\endcsname{0.00}
\expandafter\def\csname cell@movie@9@Mistral@k\endcsname{5}
\expandafter\def\csname cell@movie@9@Mistral@items\endcsname{256}
\expandafter\def\csname cell@movie@9@Mistral@flip\endcsname{0.0\%}
\expandafter\def\csname cell@movie@9@Mistral@repro\endcsname{1.00}
\expandafter\def\csname cell@movie@9@Mistral@indep\endcsname{1.00}
\expandafter\def\csname cell@movie@9@Mistral@distinct\endcsname{1}
\expandafter\def\csname cell@movie@9@Mistral@qmin\endcsname{0.73}
\expandafter\def\csname cell@movie@9@Mistral@qmax\endcsname{0.73}
\expandafter\def\csname cell@movie@9@Mistral@qrange\endcsname{0.00}
\expandafter\def\csname cell@movie@10@Mistral@k\endcsname{5}
\expandafter\def\csname cell@movie@10@Mistral@items\endcsname{2000}
\expandafter\def\csname cell@movie@10@Mistral@flip\endcsname{0.2\%}
\expandafter\def\csname cell@movie@10@Mistral@repro\endcsname{0.30}
\expandafter\def\csname cell@movie@10@Mistral@indep\endcsname{0.30}
\expandafter\def\csname cell@movie@10@Mistral@distinct\endcsname{3}
\expandafter\def\csname cell@movie@10@Mistral@qmin\endcsname{0.43}
\expandafter\def\csname cell@movie@10@Mistral@qmax\endcsname{0.43}
\expandafter\def\csname cell@movie@10@Mistral@qrange\endcsname{0.00}
\expandafter\def\csname cell@movie@5@Mistral@k\endcsname{5}
\expandafter\def\csname cell@movie@5@Mistral@items\endcsname{1560}
\expandafter\def\csname cell@movie@5@Mistral@flip\endcsname{0.1\%}
\expandafter\def\csname cell@movie@5@Mistral@repro\endcsname{1.00}
\expandafter\def\csname cell@movie@5@Mistral@indep\endcsname{1.00}
\expandafter\def\csname cell@movie@5@Mistral@distinct\endcsname{1}
\expandafter\def\csname cell@movie@5@Mistral@qmin\endcsname{1.00}
\expandafter\def\csname cell@movie@5@Mistral@qmax\endcsname{1.00}
\expandafter\def\csname cell@movie@5@Mistral@qrange\endcsname{0.00}
\expandafter\def\csname cell@movie@6@Mistral@k\endcsname{5}
\expandafter\def\csname cell@movie@6@Mistral@items\endcsname{1560}
\expandafter\def\csname cell@movie@6@Mistral@flip\endcsname{0.0\%}
\expandafter\def\csname cell@movie@6@Mistral@repro\endcsname{1.00}
\expandafter\def\csname cell@movie@6@Mistral@indep\endcsname{1.00}
\expandafter\def\csname cell@movie@6@Mistral@distinct\endcsname{1}
\expandafter\def\csname cell@movie@6@Mistral@qmin\endcsname{0.95}
\expandafter\def\csname cell@movie@6@Mistral@qmax\endcsname{0.95}
\expandafter\def\csname cell@movie@6@Mistral@qrange\endcsname{0.00}
\expandafter\def\csname cell@movie@7@Mistral@k\endcsname{5}
\expandafter\def\csname cell@movie@7@Mistral@items\endcsname{1560}
\expandafter\def\csname cell@movie@7@Mistral@flip\endcsname{0.0\%}
\expandafter\def\csname cell@movie@7@Mistral@repro\endcsname{1.00}
\expandafter\def\csname cell@movie@7@Mistral@indep\endcsname{1.00}
\expandafter\def\csname cell@movie@7@Mistral@distinct\endcsname{1}
\expandafter\def\csname cell@movie@7@Mistral@qmin\endcsname{0.48}
\expandafter\def\csname cell@movie@7@Mistral@qmax\endcsname{0.48}
\expandafter\def\csname cell@movie@7@Mistral@qrange\endcsname{0.00}
\newcommand{\permCells}{18}
\newcommand{\permBelowFive}{2}
\newcommand{\permBelowOne}{0}

\newcommand{\distMedicalTenGptoss}{12.8}

\newcommand{\distMovieNineGptoss}{10.0}

\newcommand{\distMovieTenGptoss}{0.08}

\newcommand{\tokCvMedGptoss}{0.9\%}

\newcommand{\secCvMedGptoss}{20\%}

\newcommand{\secRatioMaxLlamaR}{17.0}

\newcommand{\stackSharedRequests}{7{,}544}

\newcommand{\reasonAucDissentZero}{0.91}

\newcommand{\reasonAucDissentOne}{0.77}

\newcommand{\reasonAucDissentTwo}{0.71}

\newcommand{\reasonAucDissentThree}{0.70}

\newcommand{\countVarShareGptoss}{7\%}

\newcommand{\majWrongQwen}{17.9\%}

\newcommand{\labFlipGptoss}{9.9\%}
\newcommand{\majWrongLlamaW}{30.7\%}

\newcommand{\storeBytes}{39}
\newcommand{\storeKb}{348}
\newcommand{\dedupSaved}{6.5\%}
\newcommand{\dedupCalls}{8{,}336}
\newcommand{\sigmaGranite}{0.28}
\newcommand{\sigmaCiGranite}{[0.23, 0.33]}
\newcommand{\floorGranite}{0.26\%}
\newcommand{\smallMarginChangesGranite}{78\%}
\newcommand{\smallMarginItemsGranite}{7.8\%}

\newcommand{\sigmaLlamaR}{0.37}

\newcommand{\studyCalls}{1{,}219{,}338}
\newcommand{\studyCallsMillions}{1.2}

\newcommand{\studyLast}{2026-10-09}
\newcommand{\versionSemBench}{c814e38}
\newcommand{\versionLroBench}{6eb2930}
\newcommand{\versionUdaBench}{f2da981}
\newcommand{\versionLotus}{1.1.3}

\newcommand{\windowLro}{2026-10-02 to 2026-10-03}

\newcommand{\windowSemMain}{2026-10-02}

\newcommand{\tempItemsThree}{40}
\newcommand{\tempRepeatsThree}{20}

\newcommand{\tempZeroThree}{2.5\%}
\newcommand{\tempZeroChangedThree}{12.1\%}

\newcommand{\tempOneThree}{7.5\%}
\newcommand{\tempOneChangedThree}{41.1\%}

\newcommand{\tempDiffSeedCiThree}{[+0.7, +3.5]}

\newcommand{\tempDiffSeedCiOne}{[-7.2, -0.4]}

\newcommand{\divIdenticalPctThree}{23.5\%}

\newcommand{\divFlipRelThree}{0.18}

\newcommand{\divSameRelThree}{0.43}

\newcommand{\divRelDiffCiThree}{[-0.20, -0.02]}
\newcommand{\divLengthRhoThree}{0.57}

\newcommand{\flipGptFour}{7.5\%}

\newcommand{\notReproducedGptFour}{3}
\newcommand{\queriesGptFour}{6}

\newcommand{\flipSonnet}{4.1\%}

\newcommand{\notReproducedSonnet}{5}

\newcommand{\commercialExcluded}{Movie Q1, Medical Q1, Q4, and Q10}
\newcommand{\claimCount}{9}
\newcommand{\claimTests}{28}
\newcommand{\claimEvery}{12}

\newcommand{\claimUnstable}{16}

\newcommand{\claimSplit}{6}

\newcommand{\udaLinQueriesGptoss}{123}
\newcommand{\udaLinWithinGptoss}{70}

\newcommand{\udaLinBelowGptoss}{2}

\newcommand{\udaLinWithinSelectGptoss}{8 of 8}

\newcommand{\pubPlanDeterministicCells}{50}
\newcommand{\pubPlanDeterministicMoving}{12}

\newcommand{\pubPlanRandomizedCells}{51}
\newcommand{\pubPlanRandomizedMoving}{9}

\newcommand{\moreExecGranite}{40}

\newcommand{\moreFlipFiveGranite}{3.1\%}

\newcommand{\moreFlipMaxGranite}{5.2\%}

\newcommand{\moreLaterChangeGranite}{2.2\%}

\newcommand{\moreBracketedGranite}{74/105}

\newcommand{\moreVoteBoolGranite}{33.2\%}
\newcommand{\moreVoteTheoryGranite}{33.6\%}

\newcommand{\moreDriftCiGranite}{[-0.1, +0.1]}

\newcommand{\moreItemsLlamaW}{736}

\newcommand{\moreDriftCrossLlamaW}{1.9\%}

\newcommand{\moreDriftCiLlamaW}{[+0.9, +2.6]}

\newcommand{\lroFailEnvelopePctLlamaW}{2\%}
\newcommand{\lroFailTextPctLlamaW}{39\%}
\newcommand{\lroFailJsonPctLlamaW}{42\%}
\newcommand{\lroFailNoJsonPctLlamaW}{17\%}

\newcommand{\lroJudgeSamePairs}{2{,}506}

\newcommand{\finding}[1]{\smallskip\noindent\fbox{\parbox{\dimexpr\columnwidth-2\fboxsep-2\fboxrule}{\textbf{Finding.} #1}}\smallskip}

\setcopyright{none}
\renewcommand\footnotetextcopyrightpermission[1]{}

\title[Same Query, Different Answer]{Same Query, Different Answer: Run-to-Run Variance in Semantic Query Processing}

\author{Oktie Hassanzadeh}
\email{hassanzadeh@us.ibm.com}
\affiliation{%
  \institution{IBM}
  \city{Yorktown Heights}
  \state{NY}
  \country{USA}}
\author{Dharmashankar Subramanian}
\email{dharmash@us.ibm.com}
\affiliation{%
  \institution{IBM}
  \city{Yorktown Heights}
  \state{NY}
  \country{USA}}
\renewcommand{\shortauthors}{Hassanzadeh and Subramanian}

\begin{abstract}
Semantic query engines ask a large language model (LLM) to judge individual rows, and benchmarks compare these engines on the quality of a few executions at temperature 0.
We measure how much another execution changes.
We re-execute the text and tabular workloads of SemBench, LRO-Bench, and UDA-Bench through their own harnesses, log every one of \studyCallsMillions{} million model calls, and reanalyze SemBench's published executions.
At temperature 0, between \flipMistral{} and \flipGptoss{} of rows change their operator output over five executions, depending on the model, and up to \flipExtractGptoss{} on one task.
Plans amplify these changes exponentially in the number of rows an answer exposes, as a possible-worlds model predicts: on UDA-Bench, a few percent of extracted values change, but about two thirds of the answers do.
In SemBench's published results, \pubTextCellsMovingPct{} of text results change quality between executions, including on engines whose plans are deterministic.
Several per-query system comparisons, and \claimUnstable{} of \claimTests{} tests of LRO-Bench's implementation recommendations, are not settled by the executions the benchmarks run.
Changing rows are rows on which models disagree with each other; identical requests within one execution disagree as often as across executions; and one execution's decision margin or reasoning length predicts which rows change.
Voting reduces the variance but does not remove it, and a calibrated cascade keeps its guarantee while the rows it returns change.
We derive reporting rules for benchmarks and design rules for engines.
\end{abstract}


\begin{document}

\maketitle

% ---------------------------------------------------------------------------
\section{Introduction}
\label{sec:intro}

Semantic operators extend relational query processing with predicates and
functions that a large language model (LLM) evaluates. A semantic filter,
map, join, ranking, or aggregation sends each input row, or each pair of
rows, to the model together with a natural-language
instruction~\cite{patel2024lotus,liu2024palimpzest,shankar2024docetl,aggarwal2025cortex,dorbani2025flock}.
The filter \emph{the review is clearly positive} has no definition in the
query itself; a row qualifies if the model says it does.
Benchmarks for these systems~\cite{lao2025sembench,su2026lrobench,deng2026udabench}
report the quality, cost, and latency of each query on each system and
compare systems on these numbers.
Optimizers calibrate proxy models and cascades against the model's
outputs~\cite{kang2020proxies,zeighami2025bargain,russo2025abacus}, and
engines store semantic outputs for later reuse~\cite{dai2024uqe}.
The model's outputs thus become query answers, materialized columns, and
the labels optimizers calibrate on, so whether they repeat is a question
about the data that query processing produces.

All of these uses treat an execution of a semantic query as a property of
the system: running the query again returns the same answer.
The assumption appears safe because benchmarks execute queries at
temperature 0, which makes decoding greedy.
For LLMs in general, however, outputs at temperature 0 vary between
calls~\cite{ouyang2023chocolates,song2025greedy,atil2025nondeterminism,he2025nondeterminism,yuan2025fp32}.
A general-purpose semantic query engine cannot assume deterministic model
endpoints. Hosted APIs give no such guarantee~\cite{openai2025seed}, and
the deployment behind an endpoint changes over time; one of ours changed
between two of our campaigns (\Cref{sec:sources}). Deterministic modes of
open-source servers are opt-in, restricted to some GPUs, and cost
substantial
throughput~\cite{he2025nondeterminism,vllm2025batchinvariance,sglang2025deterministic},
and outputs still change with GPU type and count~\cite{yuan2025fp32}.
What is not known is what this variation means for query processing: how
many outputs change, whether the changes survive the operators above,
whether they change benchmark conclusions, what causes them, and what
benchmarks and engines can do.

\begin{figure}[t]
\centering
\includegraphics[width=\columnwidth]{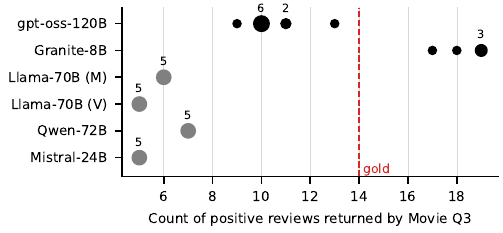}
\caption{The count that Movie Q3 returns in each execution at temperature 0, per deployment. Marker labels give the number of executions that returned the count; grey deployments returned one count in every execution.}
\Description{Dot plot of the counts returned by repeated executions of one query on six model deployments, with the gold count marked.}
\label{fig:teaser}
\end{figure}

\begin{example}
\label{ex:intro}
SemBench Movie Q3 counts the positive reviews of one film among
\cellval{movie}{3}{Gptoss}{items} candidates (\Cref{fig:teaser}).
Executed \cellval{movie}{3}{Gptoss}{k} times on gpt-oss-120B with
SemBench's own LOTUS plan and prompt at temperature 0, it returned
\cellval{movie}{3}{Gptoss}{distinct} different counts, and SemBench's
quality metric ranged from \cellval{movie}{3}{Gptoss}{qmin} to
\cellval{movie}{3}{Gptoss}{qmax}.
Llama-3.3-70B returned the same count in every execution on one inference
service (quality \cellval{movie}{3}{LlamaW}{qmin}), and the same count in
every execution on a second stack serving the same weights; but that count
was a different one (quality \cellval{movie}{3}{LlamaR}{qmin}).
\end{example}

\paragraph{This study.}
We measure run-to-run variance of semantic queries on text and tabular data
at two levels.
At the \emph{benchmark level}, we reanalyze SemBench's published results,
which contain five executions of every query on LOTUS, Palimpzest,
ThalamusDB, and BigQuery.
At the \emph{operator level}, we re-execute three benchmarks through their
own harnesses, without changing plans, prompts, or scoring, and log every
model call: SemBench's LOTUS implementations of the Movie and Medical text
queries, including three semantic joins; LRO-Bench's five LLM-enhanced
relational operators in each of their implementations; and UDA-Bench's
pipelines that extract attributes from documents and query them with SQL.
The logs give the output of every input row in every execution, which no
published benchmark records, and let us attribute each change of an answer
to changes of individual operator outputs.
We add two commercial models and a temperature control.
The central observation is that the plan, not the model alone, decides how
much an answer varies: a possible-worlds model predicts that reproduction
decays exponentially in the number of items an answer exposes, and the
measurements follow it.
An \emph{item} is one unit of work for a semantic operator: a row judged by
a filter or map, a pair judged by a join, or one value extracted from one
document. A \emph{deployment} is a model served by a particular inference
stack.

\paragraph{Contributions.}
This Experiments \& Analysis paper is an experimental analysis with a
reproducibility component. Its contributions are insights into existing
engines, operator implementations, plans, remedies, and benchmark
practices, not new methods:
\begin{itemize}
\item A possible-worlds model that separates item-level variance from
  answer reproduction and predicts, per plan operator, how plans amplify
  item changes (\Cref{sec:model}).
\item A test of published benchmark results and recommendations against
  repeated executions (\Cref{sec:published,sec:answers}).
\item Per-item logs of three benchmarks on eight deployments, with an
  analysis of how changes propagate through plans and what quality metrics
  hide (\Cref{sec:items,sec:answers}), and of which items change and why
  (\Cref{sec:sources}).
\item A replay of voting and guaranteed proxy cascades on the recorded
  outputs, and guidelines for benchmark designers and engine builders
  (\Cref{sec:remedies,sec:lessons}).
\end{itemize}

\paragraph{Summary of findings.}
\textbf{F1} (\Cref{sec:published}). About one in five published SemBench
results changes quality across executions, on every engine and also when
the plan is deterministic. Some per-query comparisons between systems need
more executions than the benchmark runs; scenario rankings are stable.
\textbf{F2} (\Cref{sec:items}). The share of items that change spans two
orders of magnitude across models, commercial ones included, and grows
with the number of executions observed. Graded judgments and free text
change more than lookups of stated facts.
\textbf{F3} (\Cref{sec:answers}). Plans amplify item changes exponentially
in the number of items an answer exposes: a few percent of changed
extractions change most SQL answers. \texttt{LIMIT} absorbs changes, and
quality metrics hide many of the rest; averaged over deployments, only
\ladderSemOut{} of SemBench queries reproduce every operator output and
\ladderSemAns{} their answer. Many of LRO-Bench's implementation
recommendations are not settled by its executions.
\textbf{F4} (\Cref{sec:sources}). Temperature 0 reduces the variance but
does not remove it. The variance belongs to each call: identical requests
within one execution disagree as often as across executions. It acts on
items close to a tie, which are items models disagree about, and a single
execution's decision margin or reasoning length identifies them.
\textbf{F5} (\Cref{sec:sources}). The deployment is part of the result:
the same model on two deployments returns different answers, variance, and
failure rates, and one deployment shifted between campaigns. None of the client settings or operator implementations we
varied removes the variance.
\textbf{F6} (\Cref{sec:remedies}). Voting reduces but does not remove
variance; re-asking only the items a first call flags gets most of its
benefit at a fraction of its cost. A cascade's statistical guarantee
survives the oracle's variance on our queries, but the rows it returns do
not. Recording operator outputs is the remedy that makes an answer
reproducible.

% ---------------------------------------------------------------------------
\section{Background and Related Work}
\label{sec:background}

\paragraph{Semantic query engines and their optimizers.}
Engines such as LOTUS~\cite{patel2024lotus}, Palimpzest~\cite{liu2024palimpzest},
DocETL~\cite{shankar2024docetl}, FlockMTL~\cite{dorbani2025flock},
Snowflake Cortex~\cite{aggarwal2025cortex}, UQE~\cite{dai2024uqe},
Sema~\cite{qi2026sema}, and iPDB~\cite{kumarasinghe2026ipdb} offer
semantic filters, maps, joins, rankings, and aggregations over tables and
documents; Evaporate~\cite{arora2023evaporate} extracts structured views
from documents.
Optimizers place semantic operators in plans~\cite{mang2026plop},
compile them into operator graphs or deterministic code~\cite{ha2026cadenza,dong2026sembaker},
order and estimate predicates~\cite{zhao2026larch,kim2026fastfilter},
choose listwise ranking and serving strategies~\cite{shin2026listk,son2026kalypso,sanmartino2026stretto},
and reduce cost with proxies and cascades, several with statistical
guarantees~\cite{kang2020proxies,zeighami2025bargain,zeighami2025fdj,zhu2026bas,chung2026proxy,russo2025abacus,liu2024llmqueries}.
All of them treat the model as an oracle with one answer per item; we
measure how far that holds and replay a guaranteed cascade when it does not.

\paragraph{Benchmarks and evaluation.}
SemBench~\cite{lao2025sembench} fixes queries, data, and ground truth for
several scenarios and implements each query natively in each system;
LRO-Bench~\cite{su2026lrobench} classifies LLM-enhanced relational operators
and compares implementations of each; UDA-Bench~\cite{deng2026udabench}
evaluates document analysis against an annotated relational view. They
report one execution or the mean of several, without per-item variance or
a test of whether their comparisons are stable.
Closest to our work, Alrashed et al.~\cite{alrashed2026stochastic} argue
that strict execution equality is the wrong criterion for evaluating
generated text-to-SQL queries with AI operators, because their results are
stochastic. We hold the query and plan fixed, as the engine benchmarks do,
measure the variance of their execution per item and per answer, trace it
to its sources, and test which published benchmark conclusions survive it.
Database benchmarking has long recognized that results depend on hidden
setup choices~\cite{raasveldt2018fair,manolescu2008repeatability}; we show
that for semantic queries the serving stack and the number of executions
are such choices.

\paragraph{Nondeterminism at temperature 0.}
Greedy decoding is deterministic only if the logits are. Inference servers
batch concurrent requests dynamically, and batch composition changes the
reduction order of floating-point kernels, so the logits of a request
depend on the requests that share its batch~\cite{he2025nondeterminism};
reduced precision amplifies the effect~\cite{yuan2025fp32}. These fixes act
inside the server, at its operator's cost; we measure what the variation
of the endpoints engines use does to query answers. Studies of code
generation~\cite{ouyang2023chocolates} and of LLM
benchmarks~\cite{song2025greedy,atil2025nondeterminism} report output
variation at temperature 0, and evaluation guidance asks for error bars over
repeated executions~\cite{miller2024errorbars,bouthillier2021variance}.
Self-consistency and semantic entropy use disagreement among samples
drawn at higher temperatures as a signal~\cite{wang2023selfconsistency,farquhar2024semantic,kadavath2022know}.
These works measure free-form
outputs or aggregate accuracy; we measure operator outputs inside query
plans, where relational operators combine many small judgments into one
answer.

\paragraph{Uncertain and crowdsourced data.}
A semantic operator whose output varies across executions defines a
distribution over relations, as in probabilistic
databases~\cite{dalvi2007probabilistic,suciu2011probabilistic,benjelloun2006uldb,jampani2008mcdb,huang2009maybms};
our propagation model (\Cref{sec:model}) is the possible-worlds semantics
of a tuple-independent database, applied to the question whether two worlds
give the same answer. Crowdsourced query
processing~\cite{franklin2011crowddb,marcus2011qurk,parameswaran2012crowdscreen}
faced noisy human judgments of rows and pairs and answered them with
redundancy, voting, and worker models~\cite{dawid1979em}; an LLM operator is
a worker that disagrees with itself, and we measure what redundancy buys.
Storing semantic outputs for reuse is a form of view
materialization~\cite{gupta1995views}, and recording which model outputs
produced an answer is a form of provenance~\cite{cheney2009provenance}.

% ---------------------------------------------------------------------------
\section{A Possible-Worlds Model and Measures}
\label{sec:model}

Consider a semantic operator applied to $n$ items. In each execution, item
$i$ receives an output $Y_i$ drawn from a distribution $p_i$ over possible
outputs. An execution is thus a possible world $Y=(Y_1,\dots,Y_n)$, and the
plan above the operator maps it to an answer $Q(Y)$.
We define the \emph{answer reproduction} of a query as the probability that
two independent executions return the same answer, $R=\Pr[Q(Y)=Q(Y')]$,
and the \emph{disagreement} of item $i$ as the probability that two
executions give it different outputs,
$\delta_i=\Pr[Y_i\ne Y'_i]=1-\sum_o p_i(o)^2$.
If items vary independently, $Y$ is a tuple-independent probabilistic
relation, and $R$ follows from the $p_i$ and the plan:
\begin{itemize}
\item If the answer determines every item's output, as for a per-row map,
  a filter returning its full result, or a join returning every match,
  then $R=\prod_i(1-\delta_i)\approx e^{-n\bar\delta}$. Reproduction decays
  exponentially in the number of items the answer exposes: with
  $\bar\delta=0.1\%$, a join over $n=1{,}560$ pairs reproduces with
  probability about $0.21$.
\item For \texttt{COUNT}, $R=\sum_c\Pr[C=c]^2$ for the Poisson-binomial count
  $C$; opposite changes cancel, so $R$ decreases only with the spread of $C$:
  for Boolean items with $\sigma^2=\sum_i p_i(1-p_i)$ large,
  $R\approx 1/(2\sigma\sqrt{\pi})$, which depends on $n$ only through
  $\sigma$.
\item Under \texttt{LIMIT}~$k$, the answer depends only on the items up to
  the $k$-th qualifying row, so $n$ is replaced by that prefix.
\end{itemize}
For other plans, such as a ranking by mean score, we evaluate $R$ by
simulating pairs of worlds through the plan.
\Cref{tab:exposure} lists, for each plan operator in our workloads, which
items decide the answer, and how often the prediction falls within 0.15
of the reproduction we measure (\Cref{sec:answers}).

\begin{table}[t]
\caption{Which items decide the answer under each plan operator, and how often the possible-worlds prediction is within 0.15 of measured reproduction: the closed form $e^{-x}$ (C) or the simulation through the plan (S). SemBench rows count query--deployment cells; the last row counts UDA-Bench query--deployment pairs.}
\label{tab:exposure}
\small
\setlength{\tabcolsep}{3pt}
\begin{tabular}{@{}llcr@{}}
\toprule
Plan operator & Items deciding the answer & & Within \\
\midrule
Full output (map, set, join) & all $n$ items & C & 23/24 \\
\texttt{LIMIT} $k$ & prefix up to the $k$-th match & S & 22/24 \\
\texttt{COUNT}, ratio, counts & spread of the count & S & 18/18 \\
\texttt{AVG} & spread of the mean & S & 6/6 \\
Rank by group mean & groups near a tie & S & 6/6 \\
SQL over extractions & values the SQL reads & C & 202/369 \\
\bottomrule
\end{tabular}
\end{table}

The model also shows why the usual measure of variance is not a property of
the operator: the share of items whose output is not identical in $k$
executions, $1-\frac{1}{n}\sum_i\sum_o p_i(o)^k$, grows with $k$.

\paragraph{Measures.}
Let $y_{ir}$ be the parsed output of item $i$ in execution $r$ of $k$; we
parse outputs exactly as the benchmark's runner does, including its default
for an unparsable reply.
The \emph{item flip rate} is the share of items whose $k$ outputs are not
all identical. Because it grows with $k$, we compare deployments at $k=5$;
for a deployment with ten executions we report the exact expectation over
all subsets of five.
The \emph{pairwise disagreement} estimates $\bar\delta$ without depending on
$k$, as the mean over items and pairs of executions of
$\mathbf{1}[y_{ir}\ne y_{is}]$.
\emph{Answer reproduction} is the share of pairs of executions whose final
answers are identical after canonical ordering; for LRO-Bench and
UDA-Bench, with hundreds of queries, we report the share of queries whose
answer is not identical in all executions.
The \emph{quality range} is the maximum minus the minimum of the
benchmark's own quality metric across executions.
The \emph{independence prediction} of $R$ draws each item from its empirical
output distribution and evaluates the plan on 2{,}000 simulated pairs of
executions.
The \emph{decision margin} is the log-probability gap between the two most
likely tokens at the first token that determines the output, taken from the
first execution.
An item is \emph{contested} if the majority outputs of the other
deployments disagree on it; because it excludes the deployment's own
output, contestedness is a proxy for the ambiguity of the item itself.
We report Wilson intervals~\cite{brown2001interval} for proportions and
percentile bootstrap intervals otherwise.

% ---------------------------------------------------------------------------
\section{Study Design}
\label{sec:design}

We ask five research questions.
\textbf{RQ1:} How much do published benchmark results change across
executions, and do system comparisons change with them?
\textbf{RQ2:} For what share of items does an operator's output change, and
how does it depend on the model and the operator?
\textbf{RQ3:} How often does the answer change, does the item level predict
it, and does the quality metric reveal it?
\textbf{RQ4:} Which items change, and which properties of the execution
environment matter?
\textbf{RQ5:} What do repetition and statistically calibrated plans buy
under this variance?

\subsection{Workloads}
\label{sec:workloads}
We restrict the study to text and tabular data and run each benchmark
through its own harness, so that prompts, operators, plans, and scoring are
the benchmark's. We list every deviation.

\paragraph{SemBench.}
For RQ1 we use SemBench's published results~\cite{lao2025sembench}: five
executions with Gemini 2.5 Flash of every query on LOTUS, Palimpzest,
ThalamusDB, and BigQuery at the scale factors of SemBench's cross-system
comparison. Headline numbers use the \pubTextQueries{} queries whose inputs
are only text or tables according to SemBench's coverage file.
For RQ2--RQ5 we execute SemBench's LOTUS runner on the Movie scenario and on
the three text queries of Medical (\Cref{tab:queries}). They cover Boolean
filters under \texttt{LIMIT}, \texttt{COUNT}, a ratio, a set result, and
\texttt{AVG}; a two-way label grouped by label; a 1--5 score per review and
per movie followed by a ranking; a 24-way disease extraction; and three
semantic self-joins.
SemBench keeps every review of the joined film at every scale factor, so the
joins (Movie Q5--Q7) require 65{,}536 comparisons each. We keep a seeded
sample of 40 of the film's 128 distinct reviews, leave every other row
unchanged, evaluate SemBench's gold SQL on the same files, and execute the
join exactly (1{,}560 pairs of distinct reviews) rather than with SemBench's
default cascade, which samples pairs at random and would add algorithmic
randomness to the model variance we measure.

% Generated by scripts/report_sembench_repeats.py. Do not edit by hand.
\begin{table}[t]
\caption{SemBench queries executed repeatedly. Items is the number of semantic-operator outputs analysed per execution; for the joins, on a 40-review sample, it excludes each review paired with itself.}
\label{tab:queries}
\small
\begin{tabular}{@{}lllr@{}}
\toprule
Query & Output & Answer & Items \\
\midrule
Movie Q1 & bool & first 5 positive reviews & 2{,}000 \\
Movie Q2 & bool & first 5 qualifying reviews & 120 \\
Movie Q3 & bool & count of positive reviews & 120 \\
Movie Q4 & bool & share of positive reviews & 120 \\
Movie Q8 & label & reviews per sentiment & 120 \\
Movie Q9 & score & score per review & 256 \\
Movie Q10 & score & movies ranked by mean score & 2{,}000 \\
Medical Q1 & bool & patients with allergy symptoms & 1{,}200 \\
Medical Q4 & bool & mean age of patients with acne & 1{,}200 \\
Medical Q10 & extract & disease per symptom text & 1{,}200 \\
\midrule
Movie Q5 & pair & first 10 same-sentiment pairs & 1{,}560 \\
Movie Q6 & pair & first 10 opposite-sentiment pairs & 1{,}560 \\
Movie Q7 & pair & all opposite-sentiment pairs & 1{,}560 \\
\bottomrule
\end{tabular}
\end{table}

\paragraph{LRO-Bench.}
LRO-Bench~\cite{su2026lrobench} defines five LLM-enhanced relational
operators (select, match, impute, cluster, order) over 27 databases, in
single- and multi-operator queries, and implements each operator with one
call for the whole operand (ALL), one call per element (ONE), or a hybrid
(SEMI). We run its pipelines unchanged on every scored query in each of the
\lroSettings{} operator--implementation settings its scripts evaluate,
without chain-of-thought prompting: \lroQueriesGptoss{} query--setting
combinations per deployment.
LRO-Bench retries a failed query and averages up to three valid executions;
we record every execution and count an unparsable reply as a failure. We
use one fixed judge, gpt-oss-120B, for impute answers so that scores are
comparable across deployments.

\paragraph{UDA-Bench.}
UDA-Bench~\cite{deng2026udabench} poses SQL queries over attributes that
must first be extracted from documents. We use its Player dataset (225
Wikipedia documents about players, teams, owners, and cities) and 132
queries covering selection, filtering, aggregation, joins, and mixtures.
Nine queries name a column differently from the ground truth and fail even
on it, which leaves \udaQueries{} queries.
We follow UDA-Bench's LOTUS plan: one extraction call per document and
attribute with the plan's instruction to keep values concise, then the
query's SQL over the extracted tables, run and scored lexically by
UDA-Bench's own code. Values are truncated to their first line and cast to
the ground-truth type, as in UDA-Bench; a rejected request yields
\texttt{NULL}. The \udaGoldEmpty{} queries with an empty ground truth count
for reproduction but not for quality.

\subsection{Models, serving, and protocol}
We call the open-weight models through an OpenAI-compatible gateway
(\Cref{tab:models}): one reasoning model, three instruction-tuned models
with 70--72B parameters, one 24B model, and one 8B model. The gateway
serves Llama-3.3-70B from a managed inference service (M) and from a vLLM
deployment (V). The two deployments share the model name and the prompt: on all
\stackSharedRequests{} requests both received, the prompt token counts are
identical. They may still differ in hardware, software, numeric precision,
and load, so a difference between them is a difference of deployments.
Through a second gateway we run two commercial models, GPT-4o and Claude
Sonnet~5, on the SemBench filters and maps. The newest commercial reasoning
models we could reach reject a temperature of 0, and older ones accept the
parameter without reporting whether it was applied, so neither can be
studied at temperature 0.
Every execution uses the benchmark's runner, plan, and prompt at
temperature 0. SemBench and UDA-Bench run on LOTUS with its cache disabled
and SemBench's default of 20 concurrent requests; LRO-Bench uses its own
client with up to ten concurrent requests and five attempts per call.
We checksum the inputs before each execution and disable SemBench's data
generation, so every execution reads identical inputs in identical order.
A wrapper around each harness's model client records a hash of the request
and the output of every call, and on LOTUS also the finish reason, token
counts, reasoning length, and, where served, the top-5 log-probabilities of
the first output tokens; it changes no request except to ask for them.
A SemBench execution with a failed call is discarded and rerun; on
UDA-Bench, a transport failure is retried for the affected document only.
We use LOTUS \versionLotus{} and the benchmark versions \versionSemBench{}
(SemBench), \versionLroBench{} (LRO-Bench), and \versionUdaBench{}
(UDA-Bench). The main executions ran on \windowSemMain{} (SemBench and
UDA-Bench) and \windowLro{} (LRO-Bench); additional executions, the
commercial models, and the temperature control ran on \studyLast{}. We
call the main executions and each later batch of executions a
\emph{campaign}. The study logged \studyCalls{} model calls.

\begin{table}[t]
\caption{Model deployments. M is a managed inference service, V a vLLM deployment, and C a commercial API. The last columns give executions per query on SemBench (S, including joins), LRO-Bench (L), and UDA-Bench (U); later executions of some deployments are used only where stated.}
\label{tab:models}
\small
\setlength{\tabcolsep}{4pt}
\begin{tabular}{@{}llccrrr@{}}
\toprule
Model & Type & Stack & Logprobs & S & L & U \\
\midrule
gpt-oss-120B & reasoning & M & no & 10 & 5 & 5 \\
Llama-3.3-70B-Instruct & instruct & M & no & 5 & 5 & 5 \\
Llama-3.3-70B-Instruct & instruct & V & yes & 5 & 5 & -- \\
Qwen2.5-72B-Instruct & instruct & V & yes & 5 & -- & -- \\
Mistral-Small-3.1-24B & instruct & M & no & 5 & -- & -- \\
Granite-3.3-8B-Instruct & instruct & V & yes & 5 & 5 & 5 \\
GPT-4o & instruct & C & yes & 5 & -- & -- \\
Claude Sonnet 5 & instruct & C & no & 5 & -- & -- \\
\bottomrule
\end{tabular}
\end{table}

% ---------------------------------------------------------------------------
\section{RQ1: Published Benchmark Results}
\label{sec:published}

% Generated by scripts/report_published_repeats.py. Do not edit by hand.
\begin{table}[t]
\caption{SemBench's published repeats: five executions of each query with Gemini 2.5 Flash, at the scale factor SemBench uses for its repeat tables. A cell is one system on one query. Moves is the share of cells whose quality differs across the five runs. Range is the largest difference between two runs of a cell. Best $-$ mean is how far the luckiest run sits above the published mean.}
\label{tab:published}
\small
\setlength{\tabcolsep}{4pt}
\begin{tabular}{@{}lrrrrr@{}}
\toprule
System & Cells & Moves & Range $\ge 0.1$ & Max range & Best $-$ mean \\
\midrule
LOTUS & 49 & 24\% & 6\% & 0.41 & 0.010 \\
Palimpzest & 64 & 16\% & 5\% & 0.30 & 0.007 \\
ThalamusDB & 47 & 19\% & 9\% & 0.97 & 0.016 \\
BigQuery & 64 & 27\% & 8\% & 1.00 & 0.035 \\
\midrule
Text queries & 101 & 21\% & 5\% & 0.44 & 0.007 \\
All queries & 224 & 21\% & 7\% & 1.00 & 0.017 \\
\bottomrule
\end{tabular}
\end{table}

\begin{figure}[t]
\centering
\includegraphics[width=\columnwidth]{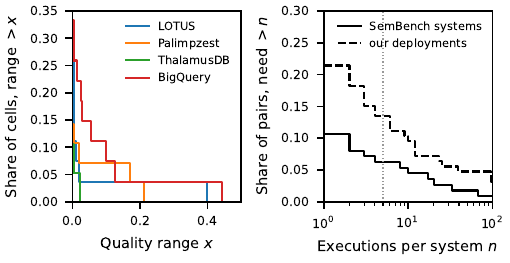}
\caption{Left: share of SemBench's published text cells whose quality range over five executions exceeds $x$. Right: share of pairs of systems on a query whose mean qualities differ and whose comparison needs more than $n$ executions per system to be separated at 95\%; dotted line at SemBench's five.}
\Description{Left: step curves of the share of cells above each quality range, per system. Right: step curves of the share of system pairs needing more than n executions, for SemBench's systems and our deployments.}
\label{fig:published}
\end{figure}

\Cref{tab:published} summarizes SemBench's five published executions per
query; we call one system on one query a \emph{cell}.
Of the \pubTextCells{} cells on text queries, \pubTextCellsMoving{}
(\pubTextCellsMovingPct{}) report a different quality in at least one
execution, \pubTextCellsBig{} by at least 0.1, and up to
\pubTextMaxRange{} (\Cref{fig:published}, left). The share ranges from \pubTextMmqaMoving{} of
\pubTextMmqaCells{} cells on MMQA to \pubTextMedicalMoving{} of
\pubTextMedicalCells{} on Medical. Including image and audio queries, the
share is the same (\pubCellsMovingPct{} of \pubCells{}), and over every
configuration SemBench publishes it is \pubAllCellsMovingPct{} of
\pubAllCells{}. Every system is affected, including BigQuery, which calls
its model inside the service.
These ranges include every source of variance, and some engines add their
own: ThalamusDB evaluates a sample of rows, Palimpzest's optimizer chooses
physical operators, and LOTUS runs semantic joins with a sampled cascade.
Engine randomness does not explain the result, however. Among cells with a
deterministic plan (LOTUS outside its joins, and BigQuery),
\pubPlanDeterministicMoving{} of \pubPlanDeterministicCells{} change
quality; among the others, \pubPlanRandomizedMoving{} of
\pubPlanRandomizedCells{} do.
In \pubEmptyCells{} cells, none on a text query, one execution returns no
rows while the others do, which indicates a failed execution recorded as a
result; excluding them, \pubCleanCellsMovingPct{} of all cells still change.

\paragraph{System comparisons.}
On text queries, \pubTextPairsReversed{} of the \pubTextPairs{} pairs of
systems whose mean quality on a query differs are ordered the other way by
at least one execution; over all queries, \pubPairsReversed{} of
\pubPairs{}, and \pubRunsReversingPct{} of all single-execution comparisons.
Five executions do not always suffice to settle a comparison either. For
\needPubMoreOne{} text pairs at least one system varies; a 95\% interval of
the difference of their means excludes zero only after a median of
\needPubMedian{} executions per system, and \needPubMoreFive{} pairs need
more than SemBench's five (\needPubMoreTen{} more than ten;
\Cref{fig:published}, right). We compute this number from the observed
standard deviations and mean difference of each pair, so it understates
the need where five executions happened to show no variance.
Scenario-level results, which average over many queries, are stable: in all
\pubScenarios{} scenarios, the best system by mean quality is best in every
execution, and the full ranking of a single execution matches the ranking
of the means in \pubOrderSame{} of \pubOrderRuns{} executions.

\finding{About one published cell in five changes quality across
executions, on every engine and also when the engine's plan is
deterministic. A per-query comparison drawn from one execution can be
reversed by another, and some need more executions than the benchmark
runs. Scenario rankings, which average over many queries, are stable, so
benchmark designers should report per-query results with their spread.}

% ---------------------------------------------------------------------------
\section{RQ2: How Many Items Change?}
\label{sec:items}

Published results contain one quality value per execution; our logs contain
every operator output. \Cref{tab:flip} reports the item flip rate per
SemBench query and deployment.

% Generated by scripts/report_sembench_repeats.py. Do not edit by hand.
\begin{table*}[t]
\caption{Repeated SemBench executions at temperature 0. F: item flip rate (\%), the share of items whose output is not identical in five executions (for gpt-oss-120B, the expected rate over five of its ten executions). R: answer reproduction, the share of pairs of executions that return an identical answer (with five executions, ten pairs, so R moves in steps of 0.1; for gpt-oss-120B, 45 pairs of ten). $\Delta$Q: range of SemBench's quality metric across executions. Shaded: the answer changes but the quality does not.}
\label{tab:flip}
\label{tab:repro}
\footnotesize
\setlength{\tabcolsep}{2.1pt}
\begin{tabular}{@{}lrrrrrrrrrrrrrrrrrr@{}}
\toprule
& \multicolumn{3}{c}{gpt-oss-120B} & \multicolumn{3}{c}{Llama-70B (M)} & \multicolumn{3}{c}{Llama-70B (V)} & \multicolumn{3}{c}{Qwen-72B} & \multicolumn{3}{c}{Mistral-24B} & \multicolumn{3}{c}{Granite-8B} \\
\cmidrule(lr){2-4}\cmidrule(lr){5-7}\cmidrule(lr){8-10}\cmidrule(lr){11-13}\cmidrule(lr){14-16}\cmidrule(lr){17-19}
Query & F & R & $\Delta$Q & F & R & $\Delta$Q & F & R & $\Delta$Q & F & R & $\Delta$Q & F & R & $\Delta$Q & F & R & $\Delta$Q \\
\midrule
Movie Q1 filter, LIMIT 5 & 4.2 & \cellcolor{black!12}.53 & .00 & 0.1 & 1 & .00 & 0.1 & 1 & .00 & 0.1 & 1 & .00 & 0.1 & 1 & .00 & 3.2 & 1 & .00 \\
Movie Q2 filter, LIMIT 5 & 4.4 & .62 & .20 & 0.0 & 1 & .00 & 0.0 & 1 & .00 & 0.0 & 1 & .00 & 0.0 & 1 & .00 & 2.5 & 1 & .00 \\
Movie Q3 filter, COUNT & 2.8 & .36 & .29 & 0.0 & 1 & .00 & 0.0 & 1 & .00 & 0.0 & 1 & .00 & 0.0 & 1 & .00 & 2.5 & .30 & .14 \\
Movie Q4 filter, ratio & 3.1 & .27 & .21 & 0.0 & 1 & .00 & 0.0 & 1 & .00 & 0.0 & 1 & .00 & 0.0 & 1 & .00 & 5.0 & .10 & .21 \\
Movie Q8 label, GROUP BY & 3.3 & .29 & .08 & 0.0 & 1 & .00 & 0.0 & 1 & .00 & 0.0 & 1 & .00 & 0.0 & 1 & .00 & 1.7 & .40 & .04 \\
Movie Q9 score per review & 19.4 & .00 & .03 & 0.0 & 1 & .00 & 0.8 & \cellcolor{black!12}.20 & .00 & 0.4 & .40 & .01 & 0.0 & 1 & .00 & 3.1 & .00 & .01 \\
Movie Q10 score, AVG, rank & 17.3 & .00 & .03 & 0.7 & \cellcolor{black!12}.10 & .00 & 0.6 & .00 & .01 & 0.1 & 1 & .00 & 0.2 & \cellcolor{black!12}.30 & .00 & 3.1 & .00 & .01 \\
Medical Q1 filter, set & 4.5 & .00 & .04 & 0.4 & .00 & .01 & 0.0 & 1 & .00 & 0.0 & 1 & .00 & 0.0 & 1 & .00 & 2.7 & .00 & .04 \\
Medical Q4 filter, AVG(age) & 0.8 & .00 & .02 & 0.1 & .60 & .01 & 0.1 & \cellcolor{black!12}.40 & .00 & 0.1 & .60 & .01 & 0.0 & 1 & .00 & 0.1 & \cellcolor{black!12}.40 & .00 \\
Medical Q10 24-way extract & 24.0 & .00 & .03 & 7.8 & .00 & .04 & 0.7 & \cellcolor{black!12}.00 & .00 & 0.0 & 1 & .00 & 0.1 & \cellcolor{black!12}.60 & .00 & 3.8 & .00 & .01 \\
\bottomrule
\end{tabular}
\end{table*}

\paragraph{The model sets the order of magnitude.}
Pooled over the ten SemBench filter and map queries, the flip rate over five
executions ranges from \flipMistral{} for Mistral-24B to \flipGptoss{}
(95\% CI \flipCiGptoss{}) for gpt-oss-120B (\Cref{tab:flip}).
Averaged over queries instead of pooled over items, the range is
\flipMacroMistral{} to \flipMacroGptoss{}. The reasoning model
changes items on every query, and
every one of its replies parses under the runner's rules.
The rate is not a fixed property of a deployment: for gpt-oss-120B it grows
from \flipKTwo{} over two executions to \flipKFive{} over five and
\flipKTen{} over ten, without levelling off (\Cref{fig:amp}, left); on 1--5
scores it grows from \flipKTwoScore{} to \flipKTenScore{}. A few executions
therefore reveal only part of the items that can change. The pairwise disagreement, which does not
depend on the number of executions, is \disagreeGptoss{} for gpt-oss-120B
and \disagreeGranite{} for Granite-8B.
On the joins (\Cref{tab:joins}), gpt-oss-120B and Granite-8B change
\cellval{movie}{7}{Gptoss}{flip} and \cellval{movie}{7}{Granite}{flip} of
the Q7 verdicts, against at most \cellval{movie}{7}{Qwen}{flip} for
Mistral-24B, Qwen-72B, and Llama-70B (V). On UDA-Bench (\Cref{tab:uda}),
gpt-oss-120B changes \udaValueFlipPctGptoss{} of \udaValuesGptoss{}
extracted values and Llama-70B (M) \udaValueFlipPctLlamaW{}.
\udaNullShareGptoss{} of gpt-oss-120B's changes are between a value and an
empty reply; counting only changes between two values, the rates are
\udaValueToValueGptoss{}, \udaValueToValueLlamaW{}, and
\udaValueToValueGranite{} for gpt-oss-120B, Llama-70B (M), and Granite-8B.

% Generated by scripts/commercial_report.py. Do not edit by hand.
\begin{table}[t]
\caption{Item flip rate over five executions, in percent, of the commercial models on the SemBench filters and maps, by output type, next to the lowest and highest rate among the six open-weight deployments. Movie Q1, Medical Q1, Q4, and Q10 are excluded for all deployments: the content filter of GPT-4o rejected every attempt.}
\label{tab:commercial}
\small
\setlength{\tabcolsep}{4pt}
\begin{tabular}{@{}lrrr@{}}
\toprule
Output & GPT-4o & Claude Sonnet 5 & Open-weight \\
\midrule
Boolean filters & 0.00 & 2.22 & 0.00--3.5 \\
Two-way label & 0.83 & 0.00 & 0.00--3.3 \\
1--5 scores & 9.04 & 4.65 & 0.13--17.5 \\
\midrule
Pooled & 7.49 & 4.13 & 0.11--15.1 \\
\bottomrule
\end{tabular}
\end{table}

\paragraph{Commercial models vary too.}
On the same filters and maps, GPT-4o changes \flipGptFour{} of items over
five executions and Claude Sonnet~5 \flipSonnet{} (\Cref{tab:commercial}).
Both rates lie inside the range of the open-weight deployments, and for
both models the 1--5 scores change most.
Their answers are not identical in all executions on
\notReproducedGptFour{} and \notReproducedSonnet{} of the
\queriesGptFour{} queries.
The content filter in front of GPT-4o rejected every attempt on
\commercialExcluded{}. On Movie Q10, it rejected some executions and
passed others with the same prompts. A filter outside the model is thus
one more source of executions that differ.

\begin{figure}[t]
\centering
\includegraphics[width=\columnwidth]{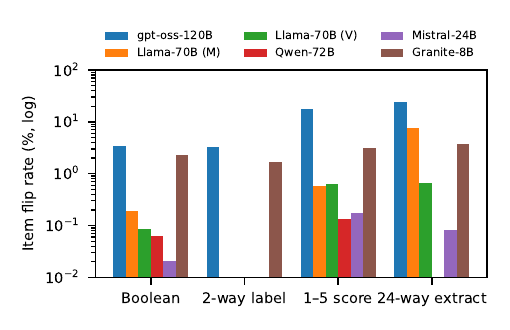}
\caption{Item flip rate over five executions by the output type of the operator, pooled over SemBench queries (log scale; zero shown at 0.01\%).}
\Description{Grouped bar chart of item flip rates per model for Boolean, label, score, and extraction outputs.}
\label{fig:kind}
\end{figure}

\paragraph{The task sets the rest.}
Within a model, the flip rate depends on what the operator asks
(\Cref{fig:kind}). For gpt-oss-120B it is \flipBoolGptoss{} on Boolean
filters and \flipLabelGptoss{} on the two-way label. On 1--5 scores it is
\flipScoreGptoss{}, and on the 24-way diagnosis \flipExtractGptoss{}.
Llama-70B (M) flips \flipBoolLlamaW{} of Boolean outputs but
\flipExtractLlamaW{} of extractions. A model that appears
deterministic on filters can vary considerably on maps.
The size of the output space does not explain this on its own. UDA-Bench's
extractions have open output spaces, yet most values are stable; the
attributes that change most are those for which a document mentions several
candidates, such as a team's owners or a city's population.
In a separate control, five executions of gpt-oss-120B on thirteen filter
and map tasks of our own over 100 synthetic product reviews, copying a
city never changed, whereas a six-way sentiment
score changed on \prodFlipSentimentScore{} of reviews and a free-text
summary on \prodFlipSummary{}.

\finding{At temperature 0, the share of items whose output changes over five
executions ranges from \flipMistral{} to \flipGptoss{}, and to
\flipExtractGptoss{} on one query, and keeps growing with more executions.
The model sets the order of magnitude; within a model, graded judgments and
free text change more than lookups of stated facts, whatever the size of the
output space.}

% ---------------------------------------------------------------------------
\section{RQ3: From Items to Answers}
\label{sec:answers}

A changed item matters only if it changes the answer. We measure answer
reproduction on SemBench's filters and maps, its joins, LRO-Bench's
operators, and SQL over UDA-Bench's extractions.

\subsection{SemBench filters and maps}

\begin{figure*}[t]
\centering
\includegraphics[width=0.86\textwidth]{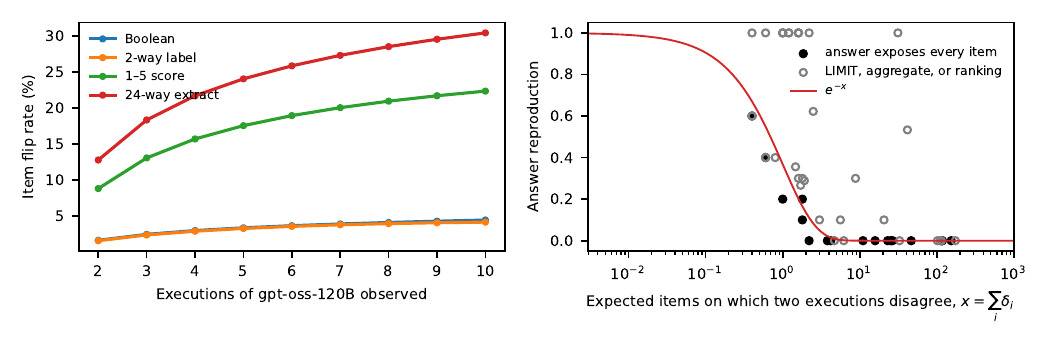}
\caption{Left: item flip rate of gpt-oss-120B against the number of executions observed, by output type (exact expectation over subsets of the ten executions). Right: answer reproduction against the expected number of items on which two executions disagree, one point per SemBench query (joins included) and deployment. Filled points are answers that expose every item, for which the possible-worlds model predicts $R\approx e^{-x}$; hollow points are answers under \texttt{LIMIT}, an aggregate, or a ranking.}
\Description{Left: line plot of flip rate rising with the number of executions for four output types. Right: scatter of reproduction against expected disagreements on a log axis, with an exponential decay curve.}
\label{fig:amp}
\end{figure*}

\paragraph{The plan sets the effect of a change.}
Answers that depend on every item, such as a per-review score or a set of
patients, change whenever any item does (\Cref{tab:flip}). On none of the
\queriesGptoss{} queries does gpt-oss-120B return the same answer in every
pair of executions.
Rankings are fragile: even Mistral-24B, which flips
\cellval{movie}{10}{Mistral}{flip} of the Movie Q10 items, reproduces its
ranking of movies in only \cellval{movie}{10}{Mistral}{repro} of execution
pairs, because movies with nearly equal means swap on a single change.

\paragraph{Item distributions predict answer stability, with a caveat.}
\Cref{fig:amp} (right) places every SemBench cell by $x=\sum_i\delta_i$,
the number of items on which two executions are expected to disagree;
cells in which no item changes have $x=0$ and are omitted. For
the \ampEveryCells{} cells whose answer exposes every item, the closed form
$e^{-x}$ alone is within 0.15 of measured reproduction in \ampEveryWithin{}
cases (mean error \ampEveryMae{}): one expected disagreement already cuts
reproduction to about a third, and no such answer with three or more was
reproduced in any pair of executions.
\ampOtherAbove{} of the other \ampOtherCells{} cells lie on or above the
curve, because \texttt{LIMIT}, aggregates, and rankings absorb some changes.
For these plans the full independence prediction, which accounts for the
plan, is close to measured reproduction for every SemBench plan operator
(\Cref{tab:exposure}).
Nor do changes concentrate in one anomalous execution: in every cell with at
least ten deviations from the item's majority output, the execution with
the most deviations accounts for at most \topRunShare{} of them. A
permutation test of the execution labels of each item agrees: of
\permCells{} such cells, \permBelowFive{} have $p<0.05$ and \permBelowOne{}
have $p<0.01$, as expected if executions are independent.
Because this in-sample fit uses the same executions it predicts, we also
fit the item distributions of gpt-oss-120B on its first five executions and
predict reproduction among the other five. The prediction is within 0.15 on
\heldoutWithin{} of \heldoutQueries{} queries. On Movie Q2, the first five
executions show no change among the rows that decide its \texttt{LIMIT}, so
the model predicts \heldoutWorstPred{}, but the later five reproduce in only
\heldoutWorstMeas{} of pairs: rare changes take more executions to observe
than to matter.

\paragraph{Quality metrics hide answer changes.}
In the shaded cells of \Cref{tab:flip}, the answer changes while quality
does not. On Movie Q1, gpt-oss-120B scores
\cellval{movie}{1}{Gptoss}{qmin} in all \cellval{movie}{1}{Gptoss}{k}
executions, yet only \cellval{movie}{1}{Gptoss}{repro} of execution pairs
return the same five reviews, because any five qualifying reviews earn full
credit. When the metric does react, the effect can be large: the quality of
Movie Q3 on gpt-oss-120B ranges from \cellval{movie}{3}{Gptoss}{qmin} to
\cellval{movie}{3}{Gptoss}{qmax}.
Changed answers are usually close to each other. On gpt-oss-120B, two
executions of Movie Q9 differ in \distMovieNineGptoss{}\% of the review
scores, and two executions of Medical Q10 in \distMedicalTenGptoss{}\% of
the diagnoses, on average; the movie rankings of Q10 differ by
$1-\tau_b=\distMovieTenGptoss{}$.
Treating the six deployments as the systems of a benchmark,
\modelPairsReversed{} of the \modelPairs{} deployment pairs whose mean
quality on a query differs are reversed by some pair of single executions,
and \needDepMoreFive{} would need more than five executions per deployment
to be separated at 95\% (\Cref{fig:published}, right, dashed).

% Generated by scripts/report_text_suites.py. Do not edit by hand.
\begin{table*}[t]
\caption{LRO-Bench, five executions of every scored query in each implementation. Per deployment: queries with a failed execution, answered queries whose answer differs between executions, and the mean range of LRO-Bench's quality metric over answered queries. ALL issues one call for the operand, ONE one call per element, SEMI a hybrid.}
\label{tab:lro}
\small
\setlength{\tabcolsep}{3pt}
\begin{tabular}{@{}llrrrrrrrrrrrr@{}}
\toprule
& & \multicolumn{3}{c}{gpt-oss-120B (M)} & \multicolumn{3}{c}{Llama-70B (V)} & \multicolumn{3}{c}{Llama-70B (M)} & \multicolumn{3}{c}{Granite-8B (V)} \\
\cmidrule(lr){3-5}\cmidrule(lr){6-8}\cmidrule(lr){9-11}\cmidrule(lr){12-14}
Operator & Impl. & Fail & Differ & Range & Fail & Differ & Range & Fail & Differ & Range & Fail & Differ & Range \\
\midrule
Select & ALL & 0 & 21/60 & 0.056 & 0 & 7/60 & 0.008 & 27 & 7/33 & 0.049 & 12 & 14/48 & 0.033 \\
 & ONE & 1 & 25/60 & 0.055 & 6 & 9/55 & 0.008 & 35 & 2/27 & 0.002 & 13 & 5/47 & 0.017 \\
\addlinespace[2pt]
Match & ALL & 0 & 30/60 & 0.145 & 3 & 9/59 & 0.021 & 41 & 11/26 & 0.089 & 17 & 9/43 & 0.050 \\
 & ONE & 9 & 33/55 & 0.074 & 20 & 19/40 & 0.013 & 58 & 2/3 & 0.077 & 12 & 29/49 & 0.023 \\
 & SEMI & 5 & 39/57 & 0.106 & 0 & 14/60 & 0.010 & 15 & 28/46 & 0.057 & 3 & 34/58 & 0.016 \\
\addlinespace[2pt]
Impute & ALL & 0 & 52/75 & 0.121 & 1 & 25/74 & 0.013 & 24 & 33/55 & 0.024 & 5 & 29/70 & 0.006 \\
 & ONE & 0 & 41/60 & 0.066 & 2 & 26/58 & 0.012 & 60 & 1/1 & 0.000 & 30 & 15/31 & 0.004 \\
\addlinespace[2pt]
Cluster & ALL & 0 & 33/60 & 0.084 & 2 & 9/59 & 0.029 & 19 & 19/47 & 0.085 & 7 & 9/53 & 0.032 \\
 & ONE & 0 & 42/60 & 0.095 & 0 & 6/60 & 0.007 & 30 & 14/30 & 0.026 & 38 & 9/25 & 0.043 \\
\addlinespace[2pt]
Order & ALL & 0 & 16/30 & 0.100 & 0 & 3/30 & 0.034 & 29 & 1/2 & 0.167 & 27 & 2/4 & 0.500 \\
 & ONE (heap) & 0 & 13/30 & 0.050 & 5 & 2/25 & 0.040 & 29 & 1/1 & 0.100 & 20 & 2/11 & 0.000 \\
 & ONE (sort) & 0 & 17/30 & 0.113 & 5 & 8/25 & 0.042 & 29 & 1/1 & 0.100 & 21 & 4/9 & 0.041 \\
 & SEMI & 0 & 29/30 & 0.244 & 13 & 4/18 & 0.011 & 30 & 0/0 & -- & 12 & 10/20 & 0.058 \\
\addlinespace[2pt]
Multi & ALL & 1 & 22/59 & 0.220 & 7 & 28/54 & 0.130 & 11 & 24/51 & 0.115 & 60 & 0/0 & -- \\
\midrule
All & & 16 & 413/726 & 0.107 & 64 & 169/677 & 0.026 & 437 & 144/323 & 0.061 & 277 & 171/468 & 0.029 \\
\bottomrule
\end{tabular}
\end{table*}

\subsection{Semantic joins}
% Generated by scripts/report_text_suites.py. Do not edit by hand.
\begin{table}[t]
\caption{Exact semantic self-joins on 40 reviews (1{,}560 pairs of distinct reviews). Flip is the share of pairs whose verdict changes across executions (\%); Repro is answer reproduction. Q5 and Q6 return the first ten same- and opposite-sentiment pairs, Q7 all opposite-sentiment pairs. Flip rates are over five executions; for gpt-oss-120B, the expected rate over five of its ten.}
\label{tab:joins}
\small
\setlength{\tabcolsep}{3.5pt}
\begin{tabular}{@{}lrrrrrrr@{}}
\toprule
& \multicolumn{2}{c}{Q5} & \multicolumn{2}{c}{Q6} & \multicolumn{3}{c}{Q7} \\
\cmidrule(lr){2-3}\cmidrule(lr){4-5}\cmidrule(l){6-8}
Deployment & Flip & Repro & Flip & Repro & Flip & Repro & F1 range \\
\midrule
Mistral-24B (M) & 0.1 & 1.00 & 0.0 & 1.00 & 0.0 & 1.00 & 0.000 \\
Qwen-72B (V) & 0.1 & 1.00 & 0.2 & 1.00 & 0.3 & 0.10 & 0.002 \\
Llama-70B (V) & 0.3 & 1.00 & 0.2 & 1.00 & 0.2 & 0.20 & 0.001 \\
Llama-70B (M) & 2.7 & 0.10 & 1.4 & 0.30 & 1.5 & 0.00 & 0.009 \\
gpt-oss-120B (M) & 13.5 & 0.00 & 15.1 & 0.00 & 15.2 & 0.00 & 0.019 \\
Granite-8B (V) & 14.8 & 0.00 & 14.8 & 0.00 & 15.6 & 0.00 & 0.027 \\
\bottomrule
\end{tabular}
\end{table}

% Generated by scripts/report_text_suites.py. Do not edit by hand.
\begin{table}[t]
\caption{UDA-Bench Player, five executions of 123 queries. Values is the share of extracted (document, attribute) values that change (\%). Per query class, the number of queries whose result table differs between executions. F1 is UDA-Bench's quality, mean over queries with a non-empty ground truth, with the mean range across executions.}
\label{tab:uda}
\small
\setlength{\tabcolsep}{2.2pt}
\begin{tabular}{@{}lrrrrrrrrr@{}}
\toprule
& & \multicolumn{6}{c}{Answers differ} & & \\
\cmidrule(lr){3-8}
Deployment & Values & Sel. & Filt. & Agg. & Join & Mix. & All & F1 & Range \\
\midrule
Queries & & 8 & 60 & 10 & 13 & 32 & 123 & & \\
\midrule
gpt-oss-120B (M) & 7.5 & 6 & 44 & 8 & 13 & 13 & 84 & 0.44 & 0.031 \\
Llama-70B (M) & 4.0 & 8 & 40 & 10 & 10 & 15 & 83 & 0.51 & 0.011 \\
Granite-8B (V) & 5.2 & 8 & 48 & 9 & 6 & 13 & 84 & 0.38 & 0.021 \\
\bottomrule
\end{tabular}
\end{table}

A join multiplies the items an answer depends on (\Cref{tab:joins}). Q7
returns every opposite-sentiment pair among 1{,}560, so a single changed
verdict changes the answer, and only Mistral-24B, which changes none,
reproduces it. Qwen-72B and Llama-70B (V) change only
\cellval{movie}{7}{Qwen}{flip} and \cellval{movie}{7}{LlamaR}{flip} of the
verdicts, yet reproduce in only \cellval{movie}{7}{Qwen}{repro} and
\cellval{movie}{7}{LlamaR}{repro} of pairs, close to the independence
predictions of \cellval{movie}{7}{Qwen}{indep} and
\cellval{movie}{7}{LlamaR}{indep}. Q5 and Q6 return only the first ten
matches and reproduce in every execution on the same deployments.
The F1 score of Q7 barely moves, whereas the precision of Q5's ten pairs
on gpt-oss-120B ranges from \cellval{movie}{5}{Gptoss}{qmin} to
\cellval{movie}{5}{Gptoss}{qmax}.

\subsection{Relational operators and their implementations}
On LRO-Bench (\Cref{tab:lro}), a query is \emph{answered} by a deployment if
at least two of its executions succeed. The answer of gpt-oss-120B differs
between executions on \lroDifferGptoss{} of its \lroAnsweredGptoss{}
answered queries (\lroDifferPctGptoss{}), and its quality changes on
\lroMovesGptoss{}. Every operator is affected: \lroDifferPctSelectGptoss{}
of answered queries for select, \lroDifferPctMatchGptoss{} for match,
\lroDifferPctImputeGptoss{} for impute, \lroDifferPctClusterGptoss{} for
cluster, \lroDifferPctOrderGptoss{} for order, and
\lroDifferPctMultiGptoss{} for multi-operator pipelines. The
instruction-tuned models change less often (\lroDifferPctLlamaR{} for
Llama-70B (V), \lroDifferPctGranite{} for Granite-8B). Llama-70B (M) fails
on most queries (\Cref{sec:sources}), so its rate covers a different set of
queries. The mean quality range is \lroRangeGptoss{} for gpt-oss-120B and
\lroRangeLlamaR{} for Llama-70B (V).

\begin{table}[t]
\caption{LRO-Bench's implementation recommendations on our executions. $\checkmark$: the 95\% interval of the mean quality difference over queries is above zero and every execution agrees; $\sim$: the interval contains zero; --: fewer than three queries.}
\label{tab:claims}
\footnotesize
\setlength{\tabcolsep}{2.5pt}
\begin{tabular}{@{}lcccc@{}}
\toprule
Recommendation & gpt-oss & Llama (M) & Llama (V) & Granite \\
\midrule
Filter, rows: ONE $>$ ALL & $\sim$ & -- & $\sim$ & $\checkmark$ \\
Filter, columns, tables: ALL $>$ ONE & $\checkmark$ & $\checkmark$ & $\checkmark$ & $\checkmark$ \\
Match, rows: SEMI $\geq$ ALL & $\sim$ & -- & $\sim$ & $\sim$ \\
Match, cells, columns: ALL $>$ SEMI & $\checkmark$ & $\sim$ & $\sim$ & $\checkmark$ \\
Impute, columns: ONE $>$ ALL & $\sim$ & -- & $\sim$ & $\sim$ \\
Impute, cells, rows: ALL $>$ ONE & $\checkmark$ & -- & $\checkmark$ & $\sim$ \\
Cluster: ALL $>$ ONE & $\checkmark$ & $\checkmark$ & $\sim$ & $\sim$ \\
Order: ALL $>$ SEMI & $\checkmark$ & -- & $\sim$ & -- \\
Order: ALL $>$ ONE (heap) & $\sim$ & -- & $\sim$ & -- \\
\bottomrule
\end{tabular}
\end{table}

\paragraph{Do LRO-Bench's recommendations survive?}
LRO-Bench recommends an implementation per operator and operand type. We
test \claimCount{} of these recommendations on each deployment
(\Cref{tab:claims}), comparing the mean quality of two implementations
over queries with a bootstrap interval, and checking each execution on its
own. Of \claimTests{} tests, \claimEvery{} hold in every execution and none
is reversed, but in \claimUnstable{} the interval contains zero. In
\claimSplit{} tests, single executions disagree on which implementation is
better, so a single execution could support either conclusion. Only the
recommendation of ALL over ONE for filters on columns and tables holds on
every deployment. Our models differ from those LRO-Bench
used, so this tests the stability of the recommendations, not the
published numbers.

\subsection{SQL over extracted tables}
UDA-Bench separates extraction from querying and shows the amplification
the model predicts (\Cref{tab:uda}). Llama-70B (M) changes
\udaValueFlipPctLlamaW{} of the extracted values, yet the result table of
\udaDifferLlamaW{} of \udaQueries{} queries (\udaDifferPctLlamaW{}) differs
between executions; for gpt-oss-120B, \udaValueFlipPctGptoss{} of values and
\udaDifferPctGptoss{} of answers change. Most queries read one attribute of
every document in a table, so one changed value among a hundred suffices.
On gpt-oss-120B, the answer changes for most aggregation queries and all
\udaDifferJoinGptoss{} join queries.
Each query names the attributes its SQL reads, so we can sum the
disagreement of every extracted value it reads into an exposure $x_q$, as
in \Cref{sec:model}. Because a filter or aggregate may absorb a change,
$e^{-x_q}$ is a lower bound on reproduction. It is within 0.15 of measured
reproduction on \udaLinWithinGptoss{} of \udaLinQueriesGptoss{} queries for
gpt-oss-120B and on every query that only selects attributes
(\udaLinWithinSelectGptoss{}); \udaLinBelowGptoss{} queries reproduce less
often than it predicts. Quality changes on \udaMovesGptoss{}
queries, but only slightly (mean range \udaRangeGptoss{}, at most
\udaMaxRangeGptoss{}): a user who reruns a query is far more likely to see a
different result table than a different score.

\subsection{Three levels of reproducibility}
\begin{figure}[t]
\centering
\includegraphics[width=\columnwidth]{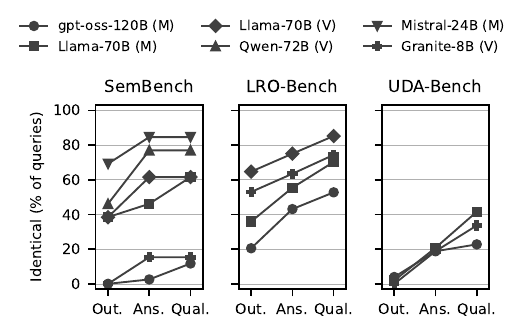}
\caption{Share of queries whose operator outputs (Out.), final answer (Ans.), and benchmark quality (Qual.) are identical over five executions, per deployment. All levels count the same queries, those with a quality score: on UDA-Bench, \udaScoredGptoss{} of \udaQueries{}; on LRO-Bench, those answered in at least two executions. LRO-Bench compares raw replies; UDA-Bench compares the extracted values each query reads.}
\Description{Slope chart with three panels, one per benchmark; each line is a deployment rising from outputs to answer to quality.}
\label{fig:ladder}
\end{figure}
\Cref{fig:ladder} summarizes the three benchmarks at three levels, over
the same scored queries and five executions. Averaged over deployments,
every operator output is identical on \ladderSemOut{} of SemBench queries,
every raw reply on \ladderLroOut{} of LRO-Bench queries (a stricter test,
since replies may differ in wording only), and every read value on
\ladderUdaOut{} of UDA-Bench queries. The answer is identical on
\ladderSemAns{}, \ladderLroAns{}, and \ladderUdaAns{}, and the quality on
\ladderSemQual{}, \ladderLroQual{}, and \ladderUdaQual{}.
Each level hides changes from the next: the gap between
outputs and answers is what plans absorb, and the gap between answers and
quality is what the metric does not see. The levels also order deployments
differently. On SemBench, Llama-70B (M) and (V) reproduce their operator
outputs equally often, but (V) reproduces more answers; on UDA-Bench, all
three deployments reproduce about a fifth of their answers, but their
quality reproduces on between \ladderUdaQualMin{} and \ladderUdaQualMax{}.
A benchmark that reports reproducibility should therefore state the level.

\finding{Whether an item change reaches the answer depends on the plan, as a
possible-worlds model predicts. Aggregates, rankings, full joins, and SQL
over extracted tables turn a few percent of changed items into changes of
most answers; \texttt{LIMIT} absorbs many changes, and quality metrics hide
many of the rest. Answer stability must be measured on answers. Item
distributions predict it per plan operator, but only once enough
executions have revealed the rare changes. Implementation recommendations
inherit the variance: many of LRO-Bench's are not settled by its
executions.}

% ---------------------------------------------------------------------------
\section{RQ4: Which Items Change, and Why}
\label{sec:sources}

\paragraph{The variance is not sampling at a hidden temperature.}
A gateway could ignore the requested temperature, or a reasoning model
could sample regardless of it. For each of Movie Q3 and Q1, we replayed the recorded requests of
\tempItemsThree{} items, half that changed in the main executions and half
that did not, \tempRepeatsThree{} times each at temperature 0, at
temperature 0 with a fixed seed, and at temperature 1, interleaved in
time. At temperature 0, the items disagree in \tempZeroThree{} of pairs of
executions; at temperature 1, in \tempOneThree{}. On the items that changed
in the main executions, the rates are \tempZeroChangedThree{} and
\tempOneChangedThree{}. Temperature 0 is therefore applied and reduces the
variance, but does not remove it. A fixed seed has no consistent effect:
the interval of its difference from temperature 0 is
\tempDiffSeedCiThree{} points on Q3 and \tempDiffSeedCiOne{} on Q1. The
gateway returns no system fingerprint that would reveal a change of
deployment.

\begin{figure*}[t]
\centering
\includegraphics[width=0.85\textwidth]{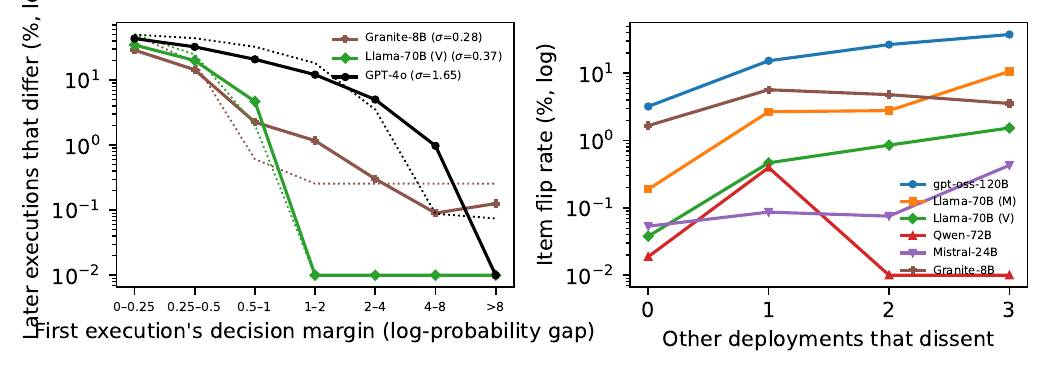}
\caption{Left: share of later executions whose output differs from the first, by the first execution's decision margin (log scale), with the fitted noise model (dotted), for deployments with log-probabilities and at least ten changing items. Right: item flip rate over five executions against the number of other deployments whose majority output differs from the plurality of the others; with five other deployments, at most three can dissent from a plurality (bins of at least 20 items; zero shown at 0.01\%).}
\Description{Left: observed share of differing later outputs falling with the decision margin, with fitted curves. Right: flip rates rising with cross-model dissent.}
\label{fig:sources}
\end{figure*}

\paragraph{Changing items are contested items.}
On SemBench, \contestedFlipped{} of changing items are contested, that is,
the majority outputs of the other deployments disagree on them, compared
with \contestedStable{} of stable items. The relation is graded
(\Cref{fig:sources}, right): gpt-oss-120B changes \dissentZeroGptoss{} of
the items on which all other deployments agree, but \dissentThreeGptoss{}
of those on which three dissent from the others' plurality; for Llama-70B
(M), the rate rises from \dissentZeroLlamaW{} to \dissentThreeLlamaW{}.
Run-to-run variance thus concentrates on items that models disagree about,
which suggests that the items themselves are ambiguous. Typical cases are
mixed reviews, such as ``Taken 3 isn't a bad time; it's often laughable
without veering totally into self-parody'' for the filter \emph{clearly
positive}, and symptoms that fit several of the 24 diseases; on each, the
other deployments disagree with one another as well.

\paragraph{Variance belongs to the call, not the execution.}
SemBench's inputs contain \dupInputs{} duplicated rows, which LOTUS sends as
separate, identical requests within one execution. If the variance came
from state that persists through an execution, the two copies would agree
within an execution more often than across executions. They do not. On gpt-oss-120B the two
copies differ in \dupWithinGptoss{} of executions (95\% CI
\dupWithinCiGptoss{}). Across executions, the pairwise disagreement is
\dupAcrossGptoss{}. On Granite-8B the two rates are \dupWithinGranite{}
and \dupAcrossGranite{}, and on the other deployments both are smaller. One Q9 review receives two different scores within
the same execution in \dupExampleDiffer{} of \dupExampleRuns{} executions, and
\dupAnyWithinGptoss{} of the \dupInputs{} duplicated inputs receive two
different outputs within at least one execution of gpt-oss-120B, so a single
answer can rest on contradictory outputs for the same input.

\paragraph{A single execution's margin predicts later changes.}
The decision margin of the first execution separates items that change in
the later executions from items that do not (\Cref{fig:sources}, left).
For Granite-8B the AUC is \marginAucGranite{} (95\% CI
\marginAucCiGranite{}, \marginFlippedGranite{} changing items).
Llama-70B (V) and Qwen-72B reach \marginAucLlamaR{} and \marginAucQwen{},
on only \marginFlippedLlamaR{} and \marginFlippedQwen{} changing items. gpt-oss-120B's stack returns no log-probabilities, but its
reasoning trace serves the same purpose: the length of the first
execution's trace predicts which items change in the other nine with AUC
\reasonAucGptoss{} (95\% CI \reasonAucCiGptoss{},
\reasonAucFlippedGptoss{} changing items). Because traces are longer on
some operators than on others, we also compute the AUC within each of the
\reasonAucQueriesGptoss{} queries with at least ten changing and ten stable
items; weighted by changing items it is \reasonAucWithinGptoss{} (from
\reasonAucWithinMinGptoss{} to \reasonAucWithinMaxGptoss{}). Nor does
contestedness explain it: within items on which zero, one, two, or three
other deployments dissent, the AUC is \reasonAucDissentZero{},
\reasonAucDissentOne{}, \reasonAucDissentTwo{}, and
\reasonAucDissentThree{}. We do not read the length as deliberation. The
temperature control above logs full traces of Movie Q3: two executions of
an item give identical traces in \divIdenticalPctThree{} of pairs, and
identical traces never give different answers. Items with longer traces
have more diverging pairs (rank correlation \divLengthRhoThree{}). Two
traces \emph{fork} at their first differing token. Pairs whose answers
differ fork earlier, at \divFlipRelThree{} of the trace's length
against \divSameRelThree{} (interval of the difference over items
\divRelDiffCiThree{}; Q1 is alike). A longer trace thus offers more
positions to fork, and an early fork changes the answer more often.
Either way, a single execution can flag the outputs to store, re-ask, or
route elsewhere.

\paragraph{One noise scale explains the margins.}
Suppose each call perturbs an item's decision margin $m$ by noise of scale
$\sigma$, so that a later execution differs from the first with
probability $\lambda+(1-\lambda)\Phi(-m/\sigma)$ for a small floor
$\lambda$. Fitted by maximum likelihood, Granite-8B gives
$\sigma=\sigmaGranite{}$ nats (95\% CI \sigmaCiGranite{}) with
$\lambda=\floorGranite{}$, and Llama-70B (V) $\sigma=\sigmaLlamaR{}$
(\Cref{fig:sources}, left). The noise is small against typical margins:
items with a margin below one nat make up \smallMarginItemsGranite{} of
Granite-8B's items but \smallMarginChangesGranite{} of its changes. Run-to-run
variance thus acts on items close to a tie, which is why it concentrates on
contested items and why voting cannot remove it (\Cref{sec:remedies}).

\paragraph{The deployment changes the answer, its variance, and whether
there is one.}
With the same model name and prompts, the majority outputs of Llama-3.3-70B on
the managed service (M) and on vLLM (V) differ on \stackDifferMovie{} of
\stackItemsMovie{} Movie items (\stackDifferPctMovie{}), whereas only
\stackFlipMovie{} items (\stackFlipPctMovie{}) change within either stack.
Each stack is nearly deterministic on Movie, but the two are
deterministically different: Movie Q3 scores
\cellval{movie}{3}{LlamaW}{qmin} on one and
\cellval{movie}{3}{LlamaR}{qmin} on the other in every execution
(\Cref{fig:teaser}). The stacks also differ in variance: on the 24-way
extraction, M flips \cellval{medical}{10}{LlamaW}{flip} of items and V
\cellval{medical}{10}{LlamaR}{flip}; on join Q5, M
\cellval{movie}{5}{LlamaW}{flip} and V \cellval{movie}{5}{LlamaR}{flip}.
On LRO-Bench, the stack decides whether there is an answer at all: at least
one execution fails for \lroFailLlamaW{} of \lroQueriesLlamaW{} queries
(\lroFailPctLlamaW{}) on M but \lroFailLlamaR{} (\lroFailPctLlamaR{}) on V.
M's replies are often not in the form LRO-Bench's code expects: of its
failed executions, \lroFailTextPctLlamaW{} precede the JSON with text,
\lroFailEnvelopePctLlamaW{} wrap it in a tool-call envelope,
\lroFailJsonPctLlamaW{} return JSON of the wrong shape, and
\lroFailNoJsonPctLlamaW{} return no JSON. A lenient parser could rescue at
most the first two groups, so M would still fail far more often than V.
These failures are reproducible: only \lroMixedFailLlamaW{} of the
\lroFailLlamaW{} failing queries succeed in another execution.
The deployment also changes over time. In the main executions, M returned
the same output in every execution for all \moreItemsLlamaW{} items of
Movie Q2--Q4, Q8, and Q9; across these and a later campaign, pairs of executions disagree
on \moreDriftCrossLlamaW{} of items (excess over pairs within a campaign
\moreDriftCiLlamaW{} points). Granite-8B on V shows no such shift
(\moreDriftCiGranite{}).

\paragraph{No operator implementation is stable.}
Across LRO-Bench's whole-operand, per-element, and hybrid implementations,
the answer of gpt-oss-120B differs on between \lroSettingMinPctGptoss{} and
\lroSettingMaxPctGptoss{} of answered queries, and no implementation is the
most stable for every operator. The implementation does change how often
the other deployments fail (\Cref{tab:lro}).

\paragraph{Client settings have no detectable effect.}
Batch composition changes the logits~\cite{he2025nondeterminism}, so we
varied one factor at a time on Movie Q3, Q8, and Q9, with ten interleaved
executions per setting (\Cref{tab:probes}): concurrency to Granite-8B,
input order, and gpt-oss-120B's reasoning effort. In all \probeSettings{} comparisons with the reference, the 95\%
bootstrap interval of the change in disagreement contains zero. The
intervals rule out only large effects, but one request at a time does not
make Granite-8B deterministic (disagreement \probeConcOneDisagreeThree{} on
Q3), consistent with a server that batches the requests of all its
tenants. Reasoning effort changes \emph{what} the model answers
but not how stably: moving from medium to low or high effort changes the
majority output of \probeEffortLowMovedNine{} and
\probeEffortHighMovedNine{} of the Q9 items, more than the within-setting
disagreement of \probeEffortMediumDisagreeNine{}.

% Generated by scripts/report_probes.py. Do not edit by hand.
\begin{table*}[t]
\caption{Cause probes on Movie Q3, Q8, and Q9. Each setting changes one factor. Disagreement is the pairwise item disagreement within the setting (\%); moved is the share of items whose majority output differs from the reference setting of the same model (\%); repeats in parentheses.}
\label{tab:probes}
\small
\begin{tabular}{@{}lrrrrrrr@{}}
\toprule
& \multicolumn{2}{c}{Movie Q3} & \multicolumn{2}{c}{Movie Q8} & \multicolumn{2}{c}{Movie Q9} & Pooled \\
\cmidrule(lr){2-3} \cmidrule(lr){4-5} \cmidrule(lr){6-7}
Setting & Disagree & Moved & Disagree & Moved & Disagree & Moved & $\Delta$ disagree [95\% CI] \\
\midrule
Granite-8B: Concurrency 1 & 1.65 (10) & 0.0 & 0.17 (10) & 0.8 & 1.68 (10) & 0.4 & -0.23 [-0.66, +0.18] \\
Granite-8B: Concurrency 8 & 1.74 (10) & 0.8 & 0.46 (10) & 0.8 & 1.55 (10) & 0.4 & -0.20 [-0.65, +0.21] \\
Granite-8B: Concurrency 20 (ref.) & 2.56 (10) & -- & 0.61 (10) & -- & 1.48 (10) & -- & -- \\
Granite-8B: Concurrency 64 & 1.85 (10) & 0.0 & 0.17 (10) & 0.8 & 1.52 (10) & 0.0 & -0.26 [-0.62, +0.07] \\
Granite-8B: Shuffled input order & 2.65 (10) & 0.8 & 0.46 (10) & 0.8 & 1.40 (10) & 0.4 & -0.06 [-0.38, +0.26] \\
\midrule
gpt-oss-120B: Reasoning effort low & 1.56 (10) & 5.0 & 0.00 (10) & 1.7 & 7.39 (10) & 14.1 & -0.97 [-2.28, +0.51] \\
gpt-oss-120B: Reasoning effort medium (ref.) & 0.63 (10) & -- & 1.87 (10) & -- & 8.82 (10) & -- & -- \\
gpt-oss-120B: Reasoning effort high & 1.74 (10) & 0.8 & 1.74 (10) & 1.7 & 7.40 (10) & 10.9 & -0.49 [-1.73, +0.65] \\
\bottomrule
\end{tabular}
\end{table*}

\finding{Changes are not uniform noise, and they are not a property of the
execution: identical requests within one execution disagree as often as
across executions. They concentrate on items that different models disagree
about, and one execution's decision margin or reasoning length predicts
them, because the variance acts on items close to a tie. Temperature 0
reduces the variance but does not remove it. The deployment changes which
answer a model gives, how much it varies, and whether it parses, and it
can shift between campaigns. None of
the settings we varied, which are client concurrency, input order,
reasoning effort, a seed, and the operator implementation, removes the
variance, so engine builders cannot configure it away.}

% ---------------------------------------------------------------------------
\section{RQ5: What Repetition and Calibration Buy}
\label{sec:remedies}

\paragraph{Voting.}
We replay repetition on the ten recorded SemBench executions of
gpt-oss-120B without new model calls, splitting them into disjoint groups
and combining each group item by item by majority (median for scores). A
vote of three equals \emph{execute twice and ask a third time only where the
two disagree}, at $2+p$ calls per item for a disagreement share $p$.
Averaged over queries, a vote of three removes \voteThreeReduction{} of the
pairwise disagreement, at most \voteThreeMaxCost{} calls per item, and a vote of five
\voteFiveReduction{} (\Cref{tab:remedies}). Neither makes the aggregate,
ranking, or Medical answers reproducible: no two combined executions return
the same answer.
The output distributions explain why. For \minorityHeavy{} of changing
items, the less common output occurs in at least three of ten executions.
With \moreExecGranite{} executions of Granite-8B on Movie Q2--Q4, Q8, and
Q9, a vote of three removes \moreVoteBoolGranite{} of the disagreement on
Boolean items, close to the \moreVoteTheoryGranite{} that independent
executions with each item's minority share predict. Voting changes item
accuracy against SemBench's labels by at most \voteMaxAccuracyChange{}
percentage points on any query, because variance is the smaller part of
the error. The majority output is wrong on between \majWrongQwen{} and
\majWrongLlamaW{} of labelled items, depending on the deployment, whereas
at most \labFlipGptoss{} of labelled items change. For the count of Movie
Q3, the variance across executions makes up \countVarShareGptoss{} of
gpt-oss-120B's squared error and nothing for most deployments; the rest is
bias.

\paragraph{Selective voting.}
Because the reasoning length of a first call predicts later changes
(\Cref{sec:sources}), an engine can re-ask only the items it flags. We
replay this on three disjoint groups of three executions on the SemBench
filters and maps: each item takes the output of the group's first
execution, except the share $q$ of items with the longest first reasoning
trace, which take the vote of all three, at $1+2q$ calls per item. Pooled
over items, a vote of three on every item
removes \selRemovedFull{} of the disagreement. Re-asking the longest-reasoning 20\% of items achieves
\selOfFullTwenty{} of that reduction at 1.4 calls per item, and 30\%
achieves \selOfFullThirty{} at 1.6; re-asking a random 20\% achieves
\selRandomOfFullTwenty{}.

% Generated by scripts/remedies_sembench_repeats.py. Do not edit by hand.
\begin{table}[t]
\caption{Repetition replayed on ten executions of gpt-oss-120B. Disagreement is the probability that two combined executions differ on an item (\%). Vote of 3 costs $2+p$ calls per item when the third call is made only where the first two disagree. Ten executions form only two groups of five, so Vote 5 rests on a single pair and can exceed Vote 3. Minority $\ge 3$ is the share of changing items whose less common output occurs in at least three of the ten executions, so that a vote of three often still returns it.}
\label{tab:remedies}
\small
\begin{tabular}{@{}lrrrrr@{}}
\toprule
Query & One & Vote 3 & Vote 5 & Cost of 3 & Minority $\ge 3$ \\
\midrule
Movie Q1 & 2.1 & 1.5 & 1.1 & 2.02 & 46\% \\
Movie Q2 & 2.1 & 1.1 & 0.8 & 2.02 & 29\% \\
Movie Q3 & 1.2 & 0.6 & 0.8 & 2.02 & 17\% \\
Movie Q4 & 1.4 & 0.6 & 0.8 & 2.02 & 17\% \\
Movie Q8 & 1.6 & 1.1 & 0.0 & 2.01 & 40\% \\
Movie Q9 & 9.9 & 5.7 & 5.5 & 2.10 & 51\% \\
Movie Q10 & 8.7 & 5.8 & 5.0 & 2.09 & 49\% \\
Medical Q1 & 2.2 & 1.4 & 1.2 & 2.02 & 43\% \\
Medical Q4 & 0.4 & 0.4 & 0.2 & 2.00 & 55\% \\
Medical Q10 & 12.8 & 8.9 & 8.4 & 2.13 & 50\% \\
\bottomrule
\end{tabular}
\end{table}

\paragraph{Calibrated cascades.}
Proxy cascades with guarantees~\cite{kang2020proxies,zeighami2025bargain}
label a sample with the expensive model, the \emph{oracle}, and choose a
threshold on a cheap proxy's score so that recall or precision with respect
to the oracle meets a target with high probability, for one oracle
labelling. We replay a recall-target cascade on Medical Q1
and Q4 (\casItems{} rows each), with Granite-8B's signed first-execution
decision margin as the proxy score and gpt-oss-120B or Llama-70B (M) as the
oracle. We calibrate for 90\% recall with 95\% confidence on 400
oracle-labelled rows, once per oracle execution, for 300 samples, and
measure recall against every oracle execution (\Cref{tab:cascade}).
On Medical Q1 with gpt-oss-120B as oracle, the same 400 rows labelled by
two different oracle executions lead to the same returned rows in only
\casReproMedOneGptoss{} of cases, and the share of calibrations that miss
the target rises from \casSameMedOneGptoss{} against the labelling
execution to \casOtherMedOneGptoss{} against another one. Both stay below
the 5\% the guarantee allows, because the slack of the confidence bound
exceeds the oracle's disagreement of \casDisMedOneGptoss{}. A precision
target was infeasible with this proxy: no threshold met it with confidence,
and the cascade returned almost no rows.

% Generated by scripts/cascade_replay.py. Do not edit by hand.
\begin{table}[t]
\caption{Proxy cascades calibrated for 90\% recall with 95\% confidence on 400 oracle-labelled rows (300 samples), with Granite-8B as proxy. Dis.: pairwise disagreement of the oracle's executions. Rows: mean rows returned. Violation: share of calibrations whose recall falls below 90\% against the oracle execution that labelled the sample (same) or another one (other). Repro: for a fixed sample, the share of pairs of oracle executions whose calibrations return the same rows. All rates in \%.}
\label{tab:cascade}
\small
\setlength{\tabcolsep}{3pt}
\begin{tabular}{@{}llrrrrr@{}}
\toprule
& & & & \multicolumn{2}{c}{Violation} & \\
\cmidrule(lr){5-6}
Query & Oracle & Dis. & Rows & Same & Other & Repro \\
\midrule
Medical Q1 & gpt-oss-120B & 2.21 & 321 & 1.0 & 2.2 & 24 \\
Medical Q1 & Llama-70B (M) & 0.18 & 350 & 1.1 & 1.1 & 84 \\
Medical Q4 & gpt-oss-120B & 0.39 & 1065 & 0.1 & 0.2 & 94 \\
Medical Q4 & Llama-70B (M) & 0.03 & 836 & 0.9 & 0.9 & 98 \\
\bottomrule
\end{tabular}
\end{table}

\paragraph{Answers with uncertainty.}
Instead of one answer, an engine could return what the possible-worlds
model implies about all of them~\cite{dalvi2007probabilistic}, fitted on
the first five executions of gpt-oss-120B and tested on the other five.
For the count of Movie Q3, the central 95\% interval [\bndLow, \bndHigh]
contains every later count but not the gold count of \bndGold{}: it
describes another execution, not the truth. For set answers, the rows
returned by all five executions (\emph{certain}~\cite{imielinski1984incomplete,libkin2016certain}) and by any
(\emph{possible}) do not bracket later executions: on Medical Q1, each
later execution omits on average \bndMedOneMissing{} certain patients and
adds \bndMedOneOutside{} that are not even possible. Five worlds miss the
rare changes: of the Granite-8B items unanimous in five executions,
\moreLaterChangeGranite{} change in a later one, and only
\moreBracketedGranite{} later Boolean answers lie between the bounds.
Uncertain answers need item probabilities from more executions or from a
per-call signal such as the margin.

\paragraph{Recording outputs.}
The remedy that does make an answer reproducible is to record the operator
outputs it came from, as response caches do~\cite{bang2023gptcache}, and
it is cheap. A stored output takes about
\storeBytes{} bytes, so one gpt-oss-120B execution of the ten SemBench
filters and maps produces \storeKb{}~KB of outputs. Deduplicating identical
requests within an execution saves \dedupSaved{} of its \dedupCalls{}
calls and removes the contradictory outputs of \Cref{sec:sources}.
Token cost varies little between executions: completion tokens vary by a
median coefficient of variation of \tokCvMedGptoss{} across executions, while wall
time varies by a median of \secCvMedGptoss{} on gpt-oss-120B, and the
slowest execution of a query takes up to \secRatioMaxLlamaR{} times as long
as the fastest on Llama-70B (V).

\finding{Repetition reduces variance but cannot remove it, because many
changing items are close to a tie. Re-asking the items with the longest
first reasoning trace achieves most of a full vote's reduction at a
fraction of its calls. A calibrated cascade keeps its guarantee
under the oracle's variance on these queries, but which rows it returns
depends on which oracle execution labelled its sample. Most error against
gold is bias, not variance. An answer is reproducible only if the operator
outputs that produced it are recorded, which costs tens of bytes per
output.}

% ---------------------------------------------------------------------------
\section{Lessons}
\label{sec:lessons}

\paragraph{For benchmark designers.}
(1)~Report per-query quality with its spread over several executions, and
choose the number of executions per comparison from the observed variance;
five did not suffice for \needPubMoreFive{} of \needPubPairs{} published
text comparisons.
(2)~Report answer reproduction next to quality.
(3)~Report flip rates at a stated number of executions, or pairwise
disagreement.
(4)~Do not report the best of several executions, and report failed
executions and their rate as a result rather than retrying them away.
(5)~Record per-item outputs, so that others can attribute variance and
replay remedies without model calls.
(6)~Report the deployment of every open-weight model.
(7)~State a recommendation between methods only if it holds across
executions, not only in one; \claimSplit{} of our tests of LRO-Bench's
recommendations would go either way on a single execution.

\paragraph{For engine builders.}
(1)~Treat a stored semantic column as one sample from a distribution, and
store the decision margin or the reasoning length with it. An engine that
wants one answer per input must deduplicate calls within an execution, not
only across executions.
(2)~Estimate a plan's answer stability from item disagreement and the
plan's shape: answers that expose $n$ items reproduce with probability
about $e^{-n\bar\delta}$, so full joins and per-row outputs need far lower
item disagreement than \texttt{LIMIT} queries.
(3)~Optimizers that calibrate on one oracle execution inherit its variance:
the guarantee may survive, but the plan does not. Record the calibration
labels with the plan.
(4)~Choose operator implementations for cost and robustness to reply
formats, not for stability.
(5)~To reproduce an answer, reuse its recorded operator outputs. Voting at
about twice the cost removes \voteThreeReduction{} of the disagreement
(averaged over queries), not all of it; if repetition is needed, spend it
on the items whose margin or reasoning length flags them.

% ---------------------------------------------------------------------------
\section{Threats to Validity}
\label{sec:threats}

\paragraph{Models, serving, and engines.}
We could not run Gemini 2.5 Flash, so RQ2--RQ5 rest on six open-weight
deployments and, for the SemBench filters and maps, two commercial models;
RQ1 covers Gemini 2.5 Flash and four engines through SemBench's own
results. No reachable model runs with and without reasoning at
temperature 0, so we cannot separate reasoning from the model.
Per-item logs need a hook into an engine's model client, which we built
for LOTUS and LRO-Bench's client; other engines enter only through RQ1. We do not control deployment load; probe and control settings are
interleaved in time. The deployment effect rests on one model
served by two deployments, the only pair available to us.
Decision margins exist only for deployments that return log-probabilities.

\paragraph{Workloads and replays.}
The study covers text and tabular data only. The joins use a 40-review
sample with an exact plan, and LRO-Bench and UDA-Bench run with the
deviations of \Cref{sec:workloads} on one of UDA-Bench's datasets.
LRO-Bench's impute scores come from a model judge that may itself vary;
in all \lroJudgeSamePairs{} pairs of executions with identical answers,
it gave identical scores.
The cascade replay fixes the proxy's scores to one execution and so
understates the variance of a real cascade.

\paragraph{Number of executions.}
Five executions bound the smallest flip rates we can detect and give answer
reproduction a granularity of 0.1, so we report intervals. Flip rates
keep growing:
on five Movie queries, Granite-8B flips \moreFlipFiveGranite{} of items
at five executions and \moreFlipMaxGranite{} at \moreExecGranite{}.

% ---------------------------------------------------------------------------
\section{Conclusion}
\label{sec:conclusion}

Temperature 0 does not make a semantic query reproducible. Item outputs
change at rates set by the model and the operator, plans amplify the
changes exponentially in the items an answer exposes, and the changes
unsettle per-query comparisons and implementation recommendations.
Benchmarks should report quality spreads,
answer reproduction, and failure rates; engines that need stable answers
should record the operator outputs that produced them.

\section*{Artifact Availability}
The drivers, the per-call logs of every execution, and the analysis scripts
that produce every number, table, and figure in this paper are available at
\url{https://github.com/semviews/sqp-variance-study}.

\bibliographystyle{ACM-Reference-Format}
\bibliography{refs}

\end{document}